\documentclass[prx,twocolumn,amsmath,amssymb,floatfix,footinbib,bibnotes,longbibliography]{revtex4-1}
\pdfoutput=1

\usepackage[breaklinks,colorlinks=true,citecolor=blue,linkcolor=magenta]{hyperref}

\usepackage{amsmath}
\usepackage{graphicx}
\usepackage{changepage}
\usepackage{fancyhdr}
\usepackage{verbatim}
\usepackage{amsthm, amssymb}
\usepackage{floatflt}
\usepackage{float}
\usepackage{array}
\usepackage[usenames,dvipsnames]{color}

\def \ket#1{{|#1\rangle}}
\def \bra#1{{\langle#1|}}

\def \mr{\mathrm}

\newcommand {\apgt} {\ {\raise-.5ex\hbox{$\buildrel>\over\sim$}}\ }
\newcommand {\aplt} {\ {\raise-.5ex\hbox{$\buildrel<\over\sim$}}\ }

\begin{document}

\title{Transport, multifractality, and the breakdown of single-parameter scaling 
at the localization transition in quasiperiodic systems}
\author{Jagannath Sutradhar, Subroto Mukerjee, Rahul Pandit, and Sumilan Banerjee}
\affiliation{Centre for Condensed Matter Theory, Department of Physics, Indian Institute 
of Science, Bangalore 560012, India}

\date\today

\begin{abstract}

There has been a revival of interest in localization phenomena in
quasiperiodic systems with a view to examining how they
differ fundamentally from such phenomena in random systems. Motivated
by this, we study transport in the quasiperiodic, one-dimentional
($1d$) Aubry-Andre model and its generalizations to $2d$ and $3d$.  We
study the conductance of open systems, connected to leads, as well as
the Thouless conductance, which measures the response of a closed
system to boundary perturbations. We find that these conductances show
signatures of a metal-insulator transition from an insulator, with
localized states, to a metal, with extended states having (a) ballistic transport ($1d$), (b) superdiffusive transport ($2d$), or (c) diffusive transport ($3d$); precisely at the transition, the
system displays sub-diffusive critical states. We calculate the beta
function $\beta(g)=d\ln(g)/d\ln(L)$ and show that, in $1d$ and $2d$,
single-parameter scaling is unable to describe the transition.
Furthermore, the conductances show strong non-monotonic variations with
$L$ and an intricate structure of resonant peaks and subpeaks.  In $1d$
the positions of these peaks can be related precisely to the properties
of the number that characterizes the quasiperiodicity of the potential;
and the $L$-dependence of the Thouless conductance is multifractal. We
find that, as $d$ increases, this non-monotonic dependence of $g$ on
$L$ decreases and, in $3d$, our results for $\beta(g)$ are reasonably well
approximated by single-parameter scaling.  

\end{abstract}

\maketitle

The single-parameter scaling theory of Abrahams, {\it et
al.},~\cite{EAbrahams1979} has played an important part in our understanding of
Anderson localization and metal-insulator transitions in disordered systems,
e.g., non-interacting electrons in a random potential \cite{PWAnderson1958}.
Localization phenomena are, however, not only restricted to random systems, but
also occur in other systems, the most prominent examples being systems with
quasiperiodic
potentials~\cite{Aubry1980,Simon1982,Sokoloff1982,Thouless1983,Kohmoto1983,Kohmoto1983_1,Ostlund1983,Ostlund1984,Sokoloff1985}.
Recently such quasiperiodic systems have attracted a lot of attention because of
the experimental observation of many-body localization (MBL) in quasiperiodic
lattices of cold atoms~\cite{Schreiber2015}. These have brought back into focus
the need to examine the essential similarities and differences between random
and quasiperiodic systems at the level of
eigenstates~\cite{Aubry1980,Simon1982,Sokoloff1982,Thouless1983,Kohmoto1983,Kohmoto1983_1,Ostlund1983,Ostlund1984,Sokoloff1985},
dynamics~\cite{Purkayastha2017,Purkayastha2018,Varma2017}, and universality
classes of localization-delocalization transitions~\cite{Khemani2017}. It has
also been argued~\cite{Khemani2017} that quasiperiodic systems provide more
robust realizations of Many Body Localization (MBL) than their random
counterparts because the former do not have rare regions, which are locally thermal. 
Therefore, we may find a stable MBL phase in dimension $d > 1$ in a 
quasiperiodic system, but not in a random system, where the MBL phase 
may be destabilised because of such rare regions~\cite{DeRoeck2016,Potriniche2018}.

Non-interacting quasiperiodic systems exhibit delocalization-localization
transitions even in one dimension ($1d$), unlike random systems in which all
states are localized in dimensions $d=1$ and $2$ for orthogonal and unitary symmetry
classes~\cite{Evers2008}.  The simplest rationale for the absence of a metallic
(delocalized) state in low-dimensional random systems and the continuous nature
of the localization-delocalization transition in three dimensions ($3d$) is provided 
by the single-parameter-scaling theory~\cite{EAbrahams1979}, which has been proposed
originally for random systems. This theory relies on only a few general
premises: (a) there is a length($L$)-dependent, dimensionless conductance,
$g(L)=G(L)/(e^2/h)$; (b) there is a single relevant scaling variable such that
$d\log(g)/d\log(L)=\beta(g)$ depends only on $g$; (c) there is a continuous and
monotonic variation of $\beta(g)$, with well-known asymptotic behaviors for
small and large conductances. Even though the conductance $g(L)$ of a finite system
(a) fluctuates strongly and (b) is a non-self-averaging quantity
\cite{Anderson1980,Altshuler1985,Lee1985}, a large number of
numerical studies~\cite{Lee1981,Pichard1981,MacKinnon1983} have provided the
justification for the single-parameter scaling theory, at least in a weak sense
\cite{Slevin2001} for typical or average conductances
\cite{Pichard1981,MacKinnon1983,Slevin2001}. Hence, to distinguish
quasiperiodic systems from random ones, it is natural to ask whether there is a
single-parameter-scaling description of the delocalization-localization
transition in quasiperiodic systems or whether quasiperiodic systems evade one
or more of the assumptions of the scaling theory. This question is particularly
relevant now because a recent study~\cite{Devakul2017} suggests that the
delocalization-localization transition in a $3d$, self-dual, quasiperiodic
model is in the same universality class as the conventional $3d$ Anderson
transition in a random system. Hence, we might expect, na\"ively, that
single-parameter scaling holds, at least, for this class of $3d$ quasiperiodic
systems. We examine this na\"ive expectation in detail.

Some recent works~\cite{Purkayastha2017,Purkayastha2018,Varma2017} have
examined open-system transport and closed-system wave-packet dynamics in
quasiperiodic chains, described by the Aubry-Andre model~\cite{Aubry1980} and
its variants~\cite{Deng2017,Ganeshan2015}, and shown that the
delocalization-localization critical point exhibits anomalous behavior: An
initially localized wave packet spreads diffusively or superdiffusively with
time in an isolated system, whereas the conductance, at high or infinite
temperature, shows subdiffusive scaling with system size, i.e., $g\sim
L^{\alpha}$ with $\alpha<-1$, for open chains connected, at its ends, to two
infinite leads~\cite{Purkayastha2017,Purkayastha2018}. These results indicate
quasiperiodic systems have much richer transport properties, at this 
critical point, than random systems.

We carry out a systematic characterization of electronic transport in the
quasiperiodic, $1d$ Aubry-Andre model and in its $2d$ and $3d$ generalizations.
We show that there are significant deviations from the expectations based on the single-parameter-scaling theory that applies to random systems. We study the
conductance of open systems connected to leads as well as the Thouless
conductance, which is a property of a closed system. Depending on the dimension $d$, these conductances show
signatures of the insulator-metal transition from an Anderson insulator to (a) a ballistic metal in $1d$,
(b) a superdiffusive metal in $2d$, or (c) a metal with diffusive transport in $3d$. Precisely at the transition, the system displays subdiffusive critical
states. We calculate the beta function $\beta(g)=d\ln(g)/d\ln(L)$ and show
that, in $1d$ and $2d$, the single-parameter scaling is unable to describe the
transition. Moreover, the conductances show strong non-monotonic variations
with $L$ and a subtle structure of resonant peaks and subpeaks.
In $1d$, we find that (a) the positions of these peaks can be related to the properties of the irrational number that characterizes the
quasiperiodicity of the potential and (b) the $L$-dependence of the Thouless
conductance is multifractal. We find that, as $d$ increases, this non-monotonic
dependence of $g$ on $L$ weakens and, in $3d$, our results for $\beta(g)$ are
well described by single-parameter scaling.  

The remainder of this paper is organised as follows. In Sec.~\ref{sec:Model} we
describe the models and give a detailed overview of our main results.
Section~\ref{sec:Results} is devoted to the description of our results for
Thouless and Landauer conductances and beta function. In
Sec.~\ref{sec:conclusion} we discuss the implications and significance of our
results.

\section{Model and overview of results}\label{sec:Model}

We study scaling of the conductance $g$ with the system-size $L$ across the
localization-delocalization (insulator-metal) transition in the well-known $1d$ 
quasiperiodic Aubry-Andre Hamiltonian~\cite{Aubry1980}

\begin{align}
\mathcal{H}&=\sum_r (e^{i \phi}c_r^\dagger c_{r+1}+\mr{h.c})+2V\sum_r \cos(2\pi b r+\phi)c_r^\dagger c_r, \label{eq.AAmodel}
\end{align}

and its $d$-dimensional generalizations~\cite{Devakul2017} (see Appendix
\ref{app:HigherDModel}). We set to unity the hopping amplitude of
electrons, which are created by $c_r^\dagger$ on the site $r$, and we characterize 
the on-site quasiperiodic potential by its strength $V$ and an
irrational number $b$, which we choose to be a quadratic irrational,
e.g., the golden ratio conjugate $b=\Phi=(\sqrt{5}-1)/2$. The phase $\phi\in
[0,2\pi)$ induces a shift of the potential, so we use it to generate a
statistical ensemble for a fixed $b$. This model~\eqref{eq.AAmodel} and its
generalizations to $2d$ and $3d$ (Appendix~\ref{app:HigherDModel}) are all
self-dual at $V=1$. In $1d$, this self-dual point coincides with the
delocalization-localization transition between a localized insulator ($V>1$)
and a ballistic metal ($V<1$)~\cite{Aubry1980}; by contrast, in $3d$, the self-dual
point lies within a diffusive-metal phase, which separates localized and ballistic
phases. These two phases are connected by a real- and momentum-space duality, akin to
that in the $1d$ model~\cite{Devakul2017}; so, in $3d$, we expect the 
localized-to-diffusive metal and ballistic-to-diffusive metal transitions 
to be dual to each other~\cite{Devakul2017}. We carry out detailed studies of 
electrical transport in all these phases and across the transitions between them
in the $1d$ Aubry-Andre model and its generalizations to $2d$ and $3d$. 
We summarize our principal results below.

\begin{figure*}[htb!]
\centering
\includegraphics[width=0.9\textwidth]{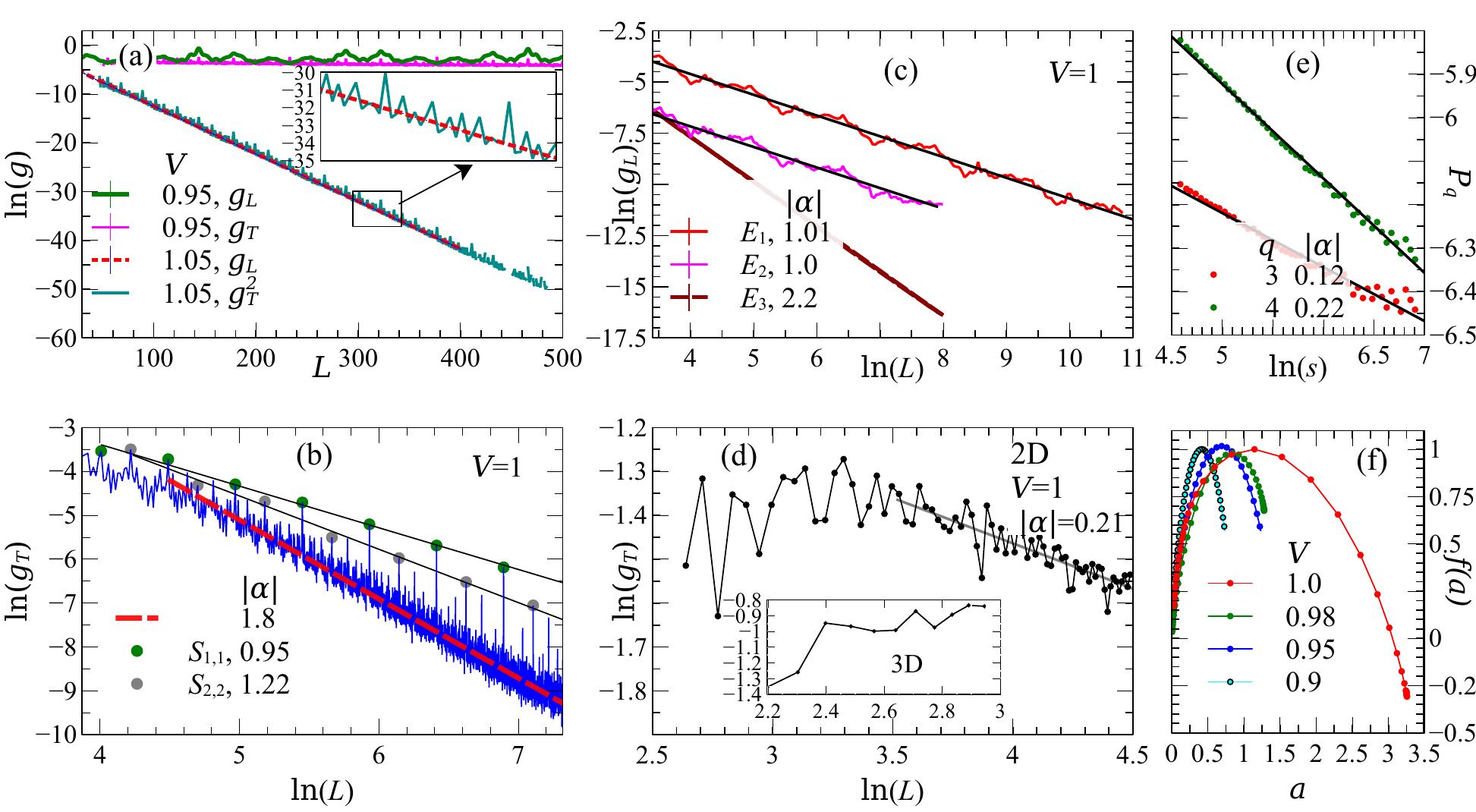}
	
\caption{{\bf Length-dependent conductances and multifractality in the
Auby-Andre model in $1d$ and its $2d$ and $3d$ generalizations}. (a)
Semilog plots versus $L$ of the conductances $g_\mr{T}$ (Thouless) and
$g_\mr{L}$ (Landauer) at illustrative values of $V$ in the metallic
($V=0.95$) and insulating ($V=1.05$) regimes in $1d$. On the metallic
side, both $g_\mr{T}$ and $g_\mr{L}$ show \textit{non-monotonic}
(roughly speaking, small-wavelength) fluctuations about an
$L$-independent mean value. On the insulating side, $g_\mr{T}^2 \propto g_\mr{L}$;
both $g_\mr{L}$ and $g_\mr{T}$ decay exponentially with $L$, the latter
only on average because $g_\mr{T}$ still displays
\textit{non-monotonic} fluctuations (enlarged view in the inset). (b)
Log-log plots versus $L$ of the Thouless conductance $g_\mr{T}$, at the
$1d$ critical point $V=1$, showing an average decay (dashed red line)
with $g_\mr{T}\propto L^{\alpha}$ and $\alpha \simeq - 1.8$, with
hierarchically organized peaks, whose heights also decay as a power of
$L$ but with different exponents (for notational simplicity denoted
generically by $\alpha$), which depend on $\mathcal{S}_{L_1,L_2}$, the
set of peaks at the lengths $L_{i+1}=L_i+L_{i-1}$, with the seed
lengths $L_1$ and $L_2$; for the illustrative sets $\mathcal{S}_{1,1}$
(green filled circles) and $\mathcal{S}_{2,2}$ (blue filled circles) we
obtain the decay exponents $\simeq -0.95$ and $\simeq -1.22$, respectively. (c) Log-log
plots versus $L$ of the Landauer conductance $g_\mr{L}$, at the $1d$
critical point $V=1$, showing an average decay $g_\mr{L}\propto
L^{\alpha(E)}$, with energy-dependent exponents $\alpha(E)$, shown for
the representative energies $E_1=0$ ($\alpha(E_1) \simeq -1.01$), 
$E_2=1.98496$ ($\alpha(E_2) \simeq -1.0$), and $E_3=0.18906032$ ($\alpha(E_3)
\simeq -2.2$; see text); note the non-monotonic fluctuations about these mean-decay
lines. (d) This non-monotonicity in log-log plots of $g_\mr{T}(L)$ versus $L$
persists in $2d$ and $3d$ (inset), as we show by illustrative data
at the the metal-insulator critical points; in $2d$, the critical
$g_\mr{T}(L)\sim L^{-0.21}$ exhibits an overall subdiffusive scaling. 
(e) A fractal analysis of the $L$-dependence of the energy-averaged 
$g_\mr{T}$, i.e., $g_\mr{T}^\infty$, versus (see
\emph{Supplementary Information}, Sec.\ref{SIsec:Multifractal} and the
main text) reveals multifractal scaling of the non-monotonic variations
of $g_\mr{T}(L)$ at the critical point. (f) A  plot of the
singularity spectrum $f(\alpha)$ versus $\alpha$ corroborates
this multifractality (see main text); note that the singularity spectrum
narrows on the metallic side $V < 1$. }

\label{fig:Conductance}
\end{figure*}
%
%

We compute the Thouless, $g_\mathrm{Th}(E,L)$, and  Landauer,
$g_\mathrm{L}(E,L)$, conductances, at a given energy $E$, for a hypercube of
volume $L^d$ ($d=1, 2,$ and $3$), as a function of the length $L$ and at zero
temperature; we obtain the averages of these conductances by varying $\phi$. We
find that \emph{even} the typical conductances, $g(L)$ (either $g=g_\mr{T}$ or
$g_\mr{L}$) are \emph{strongly} non-monotonic function of $L$; this implies that a  
strict application of single-parameter-scaling theory is untenable, especially in 
$1d$ and $2d$. This non-monotonicity is present in $3d$ too, but it is weaker
than in $2d$ and $1d$. The average $L$ dependence of these conductances, in $1d$ and
$2d$, for the localized, critical, and delocalized states, can be characterized by 
average, \emph{smooth} curves (denoted generically by $\tilde{g}(L)$); 
from these smooth curves we can obtain the associated beta functions 
$\beta(\tilde{g})$ for large system sizes. 

In $1d$, these $\beta$ functions show discontinuous jumps as we go from
localized [$\beta(\tilde{g})\sim \ln(\tilde{g})$] to ballistic metallic states
across the transition at $V=V_c=1$; the critical state exhibits sub-diffusive
power-law scaling, $\tilde{g}\sim L^{\alpha}$ such that
$\beta(\tilde{g})=\alpha < d-2 = -1$. This subdiffusive scaling is less clear
in $2d$ than in $1d$ because the onset of the scaling regime occurs only above
a large, microscopic length scale $\ell$; nevertheless, our calculation of the
open-system conductance in $2d$ suggests a similar jump in the $\beta$ function
via a sub-diffusive critical state at $V_c= 1$. Furthermore, instead of
ballistic scaling for the conductance in the metallic phase, we find
super-diffusive behavior, with a constant $\beta(\tilde{g})$ that lies between
$d-2$ and $d-1$. 

Our results in $3d$ are consistent with a continuous metal-insulator transition
at $ V_c\simeq 2.2$. We obtain scaling collapse for $g_\mr{L}(L)$ near the
transition, with a correlation-length exponent $\nu\simeq 1.6$, a value that is
close to the value of this exponent for the Anderson-localization transition in
$3d$ (as found in the recent study of Ref.~\cite{Devakul2017}, which used
moments of the wave function). Moreover, we obtain a continuous $\beta$
function from this scaling collapse; this suggests that the
single-parameter-scaling theory is a good approximation for the $3d$
quasiperiodic system we consider. However, a weak, non-monotonic $L$-dependence
of the conductance remains and indicates deviations from strict,
single-parameter scaling. We do not find a sharp transport signature of the
diffusive-metal-to-ballistic transition at $V\simeq 1/V_c$, which we expect
from duality~\cite{Devakul2017}. Given the system sizes we have been able to use 
in our study in $3d$, we find that the metallic phase, for $V\aplt 1/V_c$, exhibits 
super-diffusive scaling for $\tilde{g}_\mr{L}(L)$, with a $V$-dependent exponent 
$1<\alpha<2$ that approaches the ballistic limit ($\alpha=2$) asymptotically  
for $V\to 0$.
%
%

The non-monotonic variation of the conductance with $L$ is most prominent in
$1d$, especially for $g_\mr{T}(L)$, which exhibits resonant transport peaks for
sequences of $L$ that depend on the particular quadratic irrational number we
use; e.g., for $b=\Phi$, different sequences of peaks occur at the Fibonacci
numbers and their combinations. At the critical point, each one of these
sequences exhibits power-law scaling, i.e., $g_\mr{T}(L) \sim L^{\alpha}$ with
the exponent $\alpha$ ranging from the almost-diffusive ($\alpha \simeq -1$) to
the sub-diffusive ($\alpha<-1$) values for different sequences.  We carry out a
fractal analysis~\cite{Kantelhardt2008} of the $g_\mr{T}$ versus $L$ plot to
obtain multifractal scaling; we quantify this multifractality of the
non-monotonic variations of $g_\mr{T}$ with $L$ via the singularity spectrum
$f(\alpha)$~\cite{Kantelhardt2008}. (We use the standard notation $\alpha$ for
the crowding index~\cite{Tel2014}; this should not be confused with the
exponent $\alpha$ for the power-law scaling of the conductances.) At the
critical point, we find a broad singularity spectrum $f(\alpha)$; this narrows
in the metallic phase. Such multifractal scaling of the conductances, as a
function of $L$, is a fundamental difference between quasiperiodic and random
systems.

We show that  $g_\mr{L}(L)$ also fluctuates with $L$; however, it does not
exhibit prominent resonant peaks at distinct sequences of lengths, even in
$1d$. Hence, our results indicate a clear distinction between isolated and
open-system conductances, as measured through Thouless and Landauer
conductances, respectively.  Our results reveal very rich transport properties
for finite-size quasiperiodic systems; especially in $1d$ and $2d$, these
properties are significantly different from their counterparts random systems.

%

In the next Section we discuss our results in detail. We give some additional
aspects of our calculations and numerical computations in the \emph{Supplementary
Information}.  

\section{Results}
\label{sec:Results} \subsection{Thouless conductance} \label{sec:Thouless} 

We first characterize the response of our isolated, finite system to boundary
perturbation through the Thouless conductance, $g_\mr{T}=\delta E/\Delta_E$,
where $\delta E$ is the geometric mean of the shift of the energy levels,
around energy $E$, when we change the boundary conditions from periodic,
$\psi(r_\mu+L)=\psi(r_\mu)$, to antiperiodic, $\psi(r_\mu+L)=-\psi(r_\mu)$,
\cite{Edwards1972,Thouless1974}, in a particular direction $\mu=1,\ldots,d$;
$\Delta_E$ is the mean level spacing at energy $E$ (see \emph{Supplementary
Information}, Sec.\ref{SIsec:Thouless}).  In a diffusive metal, $g_\mr{T}$ can
be argued to be the same as the usual Landauer
$g_\mr{L}$~\cite{Edwards1972,Thouless1974,Anderson1980_1} and, in the
insulating state, it is expected that $\ln(g_\mr{L})\propto
\ln(g_\mr{T})$~\cite{Braun1997}.  However, it should be noted that $g_\mr{T}$
is a property of a closed, finite system with discrete energy eigenvalues; by
contrast, in the usual transport set up, the system is connected to infinite
leads and hence it has a continuous spectrum. As we show below for the
quasiperiodic system we consider, this makes $g_\mr{T}(L)$ significantly
different from $g_\mr{L}(L)$.

We obtain the mean $\langle g_\mr{T}\rangle (E,L)$ or typical conductance
$\exp{(\langle \ln{g_\mr{T}(E,L)}\rangle)}$ at an energy $E$ by computing
single-particle energy eigenvalues, for both periodic and antiperiodic boundary
conditions, via numerical diagonalization of the Hamiltonian in
Eq.\eqref{eq.AAmodel} or its $d$-dimensional generalizations
[Eq.\eqref{eq:model}, see  \emph{Methods}], without the phase factor in the
hopping term. The typical and mean Thouless conductances give similar results
in $1d$. The latter, $g_\mr{T}$ at $E=0$, is plotted versus $L$ for
$b=\Phi$ in Figs.\ref{fig:Conductance}, for insulating and metallic phases
[Fig.\ref{fig:Conductance}(a)] and also at the critical point $V=1$
[Fig.\ref{fig:Conductance}(b)]. We find strong non-monotonicity of
$g_\mr{T}(L)$. We first characterize its overall $L$ dependence by a
smooth least-square-fitting curve $\tilde{g}_\mr{T}(L)$, which shows 
ballistic behavior in the metallic phase, i.e., $\tilde{g}_\mr{T}$ independent
of $L$; in contrast, the conductance in the localized phase is well described
by $\tilde{g}_\mr{T}(L)\simeq g_0(V)e^{-L/\xi}$ even very close to the
transition, $V\apgt V_c$; $g_0$ denotes conductance at a microscopic length
scale $\ell \approx 1$ and varies with $V$. However, the critical state
exhibits an overall power-law dependence on $L$, $\tilde{g}_\mr{T}\sim
L^{\alpha}$ [the dashed red line in Fig.\ref{fig:Conductance}(b)] with
$\alpha\simeq -1.8$ up to the maximum system size we have studied ($L=3000$). 

The non-monotonicity of the Thouless conductance is clearly manifested in the
peak and sub-peak structure of $g_\mr{T}(L)$, in both the metallic and
insulating phases [Fig.\ref{fig:Conductance}(a)]. These peaks are most striking
at the critical point [Fig.\ref{fig:Conductance}(b)], where we find
hierarchically organized peaks, whose heights decay as a power of $L$ but with
different exponents (for notational simplicity denoted generically by
$\alpha$), which depend on $\mathcal{S}_{L_1,L_2}$, the set of peaks at the
lengths $L_{i+1}=L_i+L_{i-1}$, with the seed lengths $L_1$ and $L_2$; for the
illustrative sets $\mathcal{S}_{1,1}$ (green filled circles and $L_i = F_i$,
the Fibonacci numbers) and $\mathcal{S}_{2,2}$ (blue filled circles and $L_i =
2 F_i$) we obtain the decay exponents $\simeq -0.95$ and $\simeq -1.22$,
respectively. We can also identify similar sequences of peaks in the metallic
and insulating phases [Fig.\ref{fig:Conductance}(a)]. The development of a
quantitative theory of these peaks and their decay exponents
$\alpha_\mathcal{S}$ is an important challenge.  

%
%
%

Similar resonance peaks have been seen at high- or infinite-temperature
open-system transport~\cite{Purkayastha2017,Varma2017,Purkayastha2018}; however,
this resonance effect is much more striking in the $g_\mr{T}$ that we calculate. We find
similar resonant peaks for the energy-averaged or infinite-temperature Thouless
conductance $g_\mr{T}^{\infty}$ as well (see \emph{Supplementary Information}).
The existence of sharp resonant peaks in $g_\mr{T}(L)$, upto arbitrary large
lengths, is a special feature of $1d$ and points to markedly distinct
transport characteristic of quasiperiodic system compared to random systems in
$1d$. We find the the resonant peaks to be present in $2d$ and $3d$, albeit much less
prominently than in $1d$, as we show in Fig.\ref{fig:Conductance}(d) at the
metal-insulator transition $V=V_c$. 

\begin{figure*}[htb!]
\centering
\includegraphics[width=0.8\textwidth]{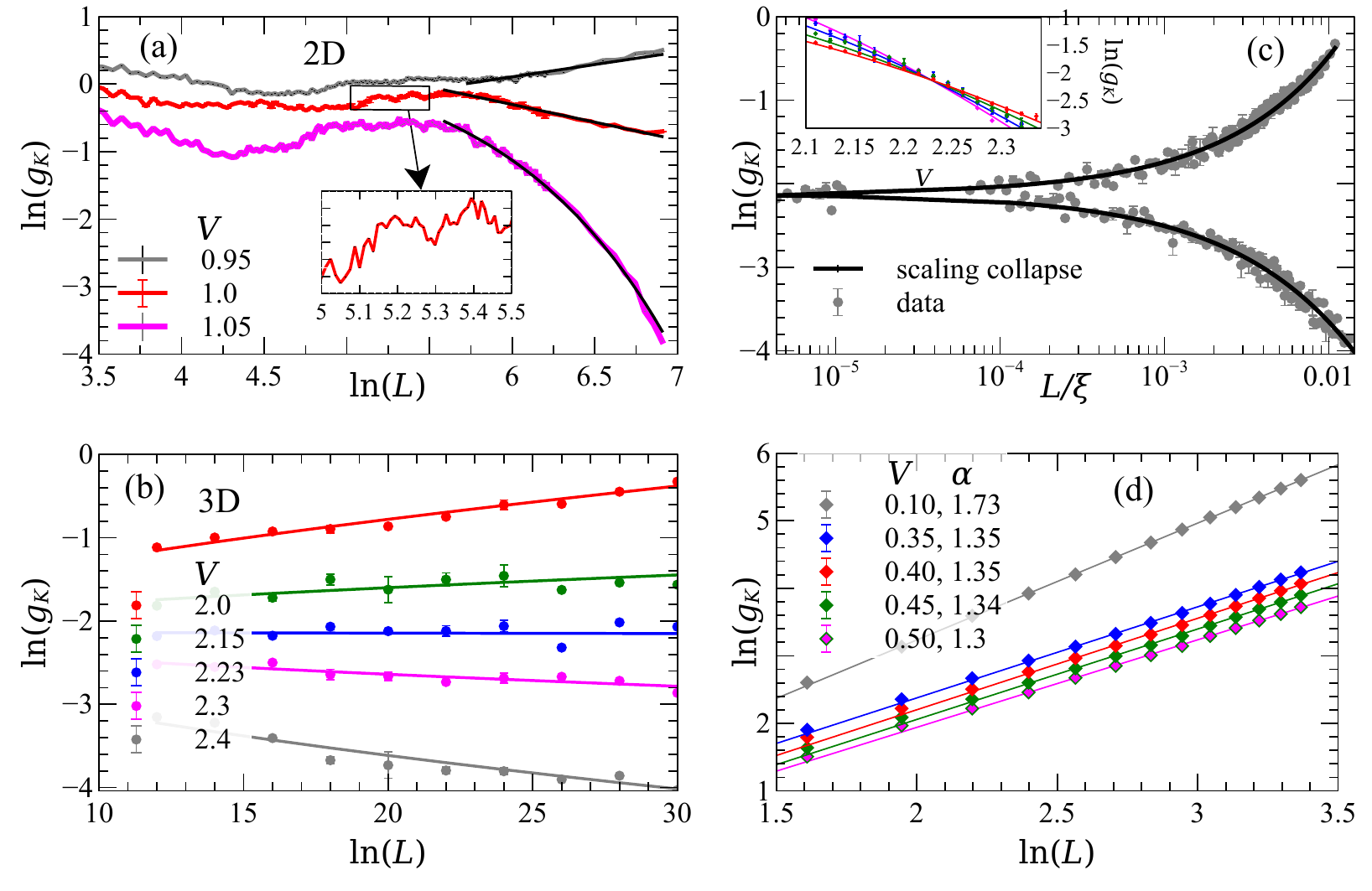}
\caption{{\bf Open-system conductance in $2d$ and $3d$.} (a) $g_\mr{K}(E=0,L)$ for insulating ($V=1.05$), critical ($V=1$) and metallic ($V=1$) states in $2d$. The solid black lines are fitting of the data for the asymptotic $L$ dependence, exponential decay for the insulating state and powerlaw scalings for critical and metallic states. The non-monotonicity of $g_\mr{K}(L)$ is shown in the inset. (b) Shows $g_\mr{K}(L)$ across localized to diffusive metal transition in $3d$. A weak non-monotonic variations, larger than the errorbars, can be seen. The solid lines are fit to the data obtained via scaling collapse, shown in (c). The inset in (c) clearly indicates the critical point at $V_c=2.22\pm 0.01$ in terms of a crossing of $g_\mr{K}$ vs. $V$ curves for different $L$. (d) Conductance near $V=1/V_c\simeq 0.45$ follows super-diffusive scaling, $g_\mr{K} \sim L^\alpha$ with $1<\alpha <2$, as shown by the fits (solid lines) to the data points and it asymptotically approaches to the ballistic scaling deep in the metallic side, e.g. $\alpha=1.73$ for $V=0.1$ as shown. }
\label{fig:Conductance_HigherD}
\end{figure*}

{\bf Conductance multifractality}:

We next ask whether the strong, non-monotonic variations of $g_\mr{T}$ with $L$
in $1d$ [Fig.\ref{fig:Conductance}(a),(b)] can be quantified in a broader
framework, rather than relying, e.g., on the number-theoretic details for
specific choices of the irrational number. Motivated by the multiple power laws
in Fig.\ref{fig:Conductance}(b) for different sequences of $L$, we carry out a
fluctuation analysis~\cite{Kantelhardt2008} of $g_\mr{T}$, as function of $L$,
by using methods that are used to treat fractal time series 
(see \emph{Supplementary Information}). We find the intriguing result thath 
$g_\mr{T}(L)$ exhibits multifractal scaling of different moments, as we
show for a few moments in \ref{fig:Conductance}(e). We also calculate the
singularity spectrum~\cite{Tel2014,Kantelhardt2008,Goldberger2000} of
$g_\mr{T}(L)$ (see \emph{Supplementary Information}). As shown in
Fig.\ref{fig:Conductance}(f), this singularity spectrum $f(\alpha)$ indicates
substantial multifractality at the critical point; it 
narrows in the metallic phase, but a substantial multifractality
still persists there. A meaningful multifractal analysis  cannot be performed in the
insulating phase because the values of the conductance become
exponentially small with $L$. We emphasize that the multifractality of
conductance reported here is distinct from usual multifractality of
wavefunction or two-point conductance~\cite{Evers2008} at the $3d$ Anderson
transition for random systems. In the latter, typical and mean conductances are
monotonic functions of $L$, and, as a result, the particular multifractality of
$g_\mr{T}(L)$, which we find here for $1d$ quasiperiodic system, would be
absent. To this end, we show in the \emph{Supplementary Information} that the
usual multifractality of the critical wavefunction, as in the $3d$
Anderson criticality \cite{Evers2008}, is also present for quasiperiodic
systems in $1d$ \cite{Dominguez1992,Mace2016,Mace2017}.

%

We have not been able to carry out a detailed multifractal analysis of 
$g_\mr{T}(L)$ in $d=2, 3$ because of the limitations of the system sizes that
we can obtain, for the Thouless conductance calculation, which require the 
numerical diagonalization of large matrices. Moreover, the scales of the
non-monotonic variations are much weaker in $2d$ and $3d$, compared to those in
$1d$, as is evident from Figs.\ref{fig:Conductance}(b),(d).

\subsection{Open-system conductance} 

We next study the conductance of open systems, starting with Aubry-Andre chain
connected to two semi-infinite leads at both ends. In this case, we compute
Landauer or Economou-Soukoulis conductance $g_\mr{L}(E)=T(E)$
\cite{Landauer1970,Economou1981}, where
$T(E)=4\sin^2k/|e^{-ik}\psi(L)-\psi(L-1)|^2$,  is the transmission coefficient
at energy $E=2t\cos{k}$, $t$ being the hopping in the tight-binding leads, and
wavefunction amplitudes $\psi(L),~\psi(L-1)$ are obtained using standard
transfer matrix method (see \emph{Supplementary Information}). For higher
dimensions $d=2,3$, we calculate the open-system conductance $g_\mr{K}$ using
Kubo formula for the system connected with leads using recursive Green function
method \cite{Lee1981,MacKinnon1983} (see \emph{Supplementary Information}). The
open-system Kubo condactance gives results identical to Landauer conductance
\cite{Fisher1981}, as we have verified for $1d$ by calculating both $g_\mr{L}$
and $g_\mr{K}$.

{\bf One dimension}:

The results for $\phi$-averaged typical conductance $\exp{\langle
\ln{g_\mr{L}(E=0,L)}\rangle}$, denoted by $g_\mr{L}$ for brevity, are plotted
in Figs.\ref{fig:Conductance}(a),(c) across metal-insulator transition in $1d$.
The overall length dependence in the metallic and insulating phases are same as
that of $g_\mr{T}(L)$, namely ballistic and localized behaviors with $L$,
respectively. However, the transport at the critical point is almost diffusive
with $g_\mr{L}\sim L^{-1.01}$ for $E=0$. Since the $1d$ Aubry-Andre chain has a
fractal energy spectra dominated by gaps
\cite{Simon1982,Kohmoto1983,Kohmoto1983_1,Ostlund1983,Ostlund1984}, it is hard
to track the $L$ dependence for an arbitrary energy as it can move into a gap
as $L$ is varied. As a result $g_\mr{L}(E)$ can cease to show the powerlaw
scaling and instead exhibit an exponentially decay with $L$ . However, the
nearly diffusive powelaw could be clearly observed till the largest system size
($L=5\times 10^4$) studied for $E=0$. For a few other energies the powelaw
could be tracked till sufficiently large $L$ as shown in
Fig.\ref{fig:Conductance}(c). Conductance at one of the energies ($E\simeq
0.189$) shows strongly subdiffusive behavior with $\alpha\simeq -2.22$. Since
conductances at diffrent energies show a range of scaling from diffusive to
subdiffusive, it is possible to obtain a overall subdiffusive conductance
scaling at higher temperature that averages over a large energy window, as in
the earlier studies \cite{Purkayastha2017,Varma2017,Purkayastha2018}. To
summarize, both $g_\mr{L}$ and $g_\mr{T}$ indicate the presence of multiple
powelaws, depending on energy and/or the sequence $\mathcal{S}$. Also, we find
the relation $g_\mr{L}\propto g_\mr{T}^2$ \cite{Anderson1980_1} to hold in the
insulating phase, however, not at the critical point, since $g_\mr{L}(E=0)$ and
$g_\mr{T}(E=0)$ follow different powelaws with $L$.

As is evident from Figs.\ref{fig:Conductance}(a),(c) (see also
Figs.\ref{fig:Disordered_landauer_cond}(a)-(d), Supplementary Information), the
Landauer conductance in $1d$ also shows strong non-monotonic dependence on $L$,
both in the metallic and critical state, even after averaging over sufficiently
large number of $\phi$'s (see \emph{Methods}) and there are peaks and subpeaks
as in $g_\mr{T}$, e.g. the dominant peaks appear at some of the Fibonacci
numbers. However, peaks are much weaker and do not appear at all $F_n$'s. The
weakening, and the absence in some cases, of the conductance peaks in
open-system conductance, as opposed to that in $g_\mr{T}$, indicate that the
leads have rather drastic effect on the system in the form of broadening and
even washing out the resonances.

\begin{figure*}[htb!]
\centering
\includegraphics[width=0.9\textwidth]{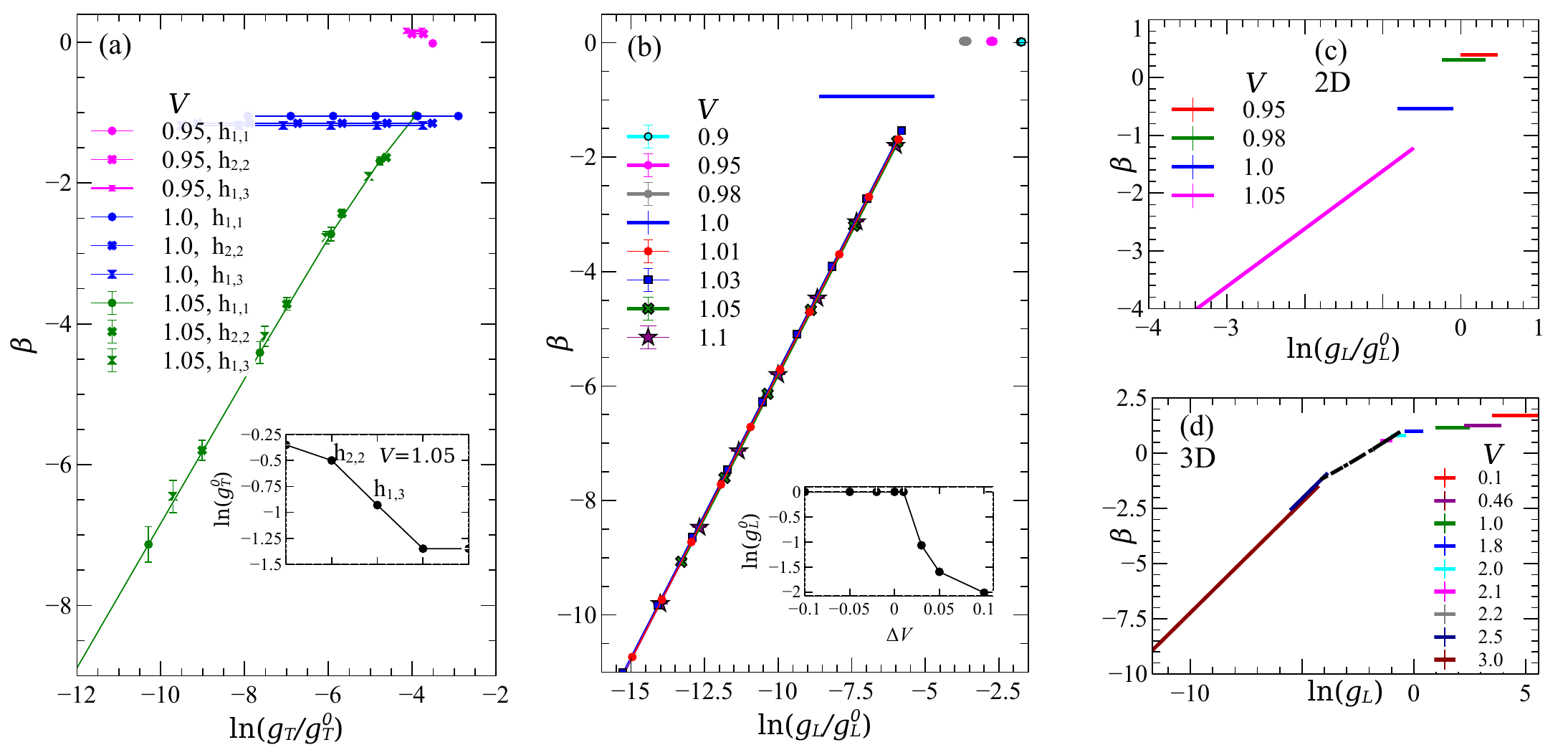}
\caption{{\bf Beta functions in $1d$, $2d$ and $3d$.} (a) $\beta(\tilde{g})$ in $1d$ extracted from $\tilde{g}_\mr{T}$ across metal-insulator transition for various sequences $\mathcal{S}_{L_1,L_2}$, obtained by fitting with exponential decay and power law for $V>1$ and $V\leq 1$, respectively. Same color with different symbols represents $\beta(\tilde{g}_\mr{T})$ calculated for different $\mathcal{S}$ and the same $V$. On the insulating side to make all the curves fall on the same line we choose different microscopic conductance ($g_\mr{T}^0$) for different sequence (inset). $\beta(\tilde{g})$ extracted in similar manner for (b) $\tilde{g}_\mr{L}$ in $1d$, (c) $\tilde{g}_\mr{K}$ in $2d$ and (d) $\tilde{g}_\mr{K}$ in $3d$ (solid lines) for values of $V$ indicated in the figure panels. In (b) and (c) the straight lines for $\beta(\tilde{g}_\mr{L})$ in the insulating side ($V>1$) has been collapsed to a single curve by choosing an appropriate $g_\mr{L}^0(V)$, as shown for $1d$ in the inset of (b). In (d) the black dashed line is the $\beta$-function calculated from the scaling collapse of Fig.\ref{fig:Conductance_HigherD}(b).}
\label{fig:BetaFunction}
\end{figure*}

{\bf Two dimensions}:

The open-system Kubo conductance $g_\mr{K}(L)$ for $E=0$ in $2d$ is shown in
Fig.\ref{fig:Conductance_HigherD}(a). Our results for system sizes up to
$1000^2$ are consistent with a metal-insulator transition at $V=V_c=1$, the
self-dual point. The conductance in the localized phase, as in $1d$, follows
$g_\mr{K}(L)\simeq g_0(V) \exp(-L/\xi)$ for $V>V_c$. The metallic phase for
$V<V_c$ is superdiffusive having $g_\mr{K}(L)\sim L^\alpha$ with
$d-2<\alpha\simeq 0.35 <d-1$, lying between diffusive and ballistic limits.
Here $g_0(V)$ is the conductance at a microscopic lenght scale $\ell$. We find
the asymptotic scaling behaviors to set in only for $L\gg \ell$, where the
microscopic length $\ell(V)$ is substantially large, varying between $L=50-500$
depending on $V$. 
ballistic increase (not shown in Fig.\ref{fig:Conductance_HigherD}(a)),
followed by an intermediate regime of length, only above which the scaling
regimes ensue.  The critical point at $V=V_c$ exhibits a subdiffusive length
scaling of conductance with $\alpha\simeq-0.52$. Again, strong non-monotonic
variations of $g_\mr{K}(L)$ is observed in all the phases, as demonstrated,
e.g., in the inset of Fig.\ref{fig:Conductance_HigherD}(a) for the critical
state.

{\bf Three dimensions}:

The results for the $3d$ conductances $g_\mr{K}(L)$ are shown in
Fig.\ref{fig:Conductance_HigherD}(b) up to $L=30$ near $V=2.2$. As evident,
non-monotonic variations of $g_\mr{K}(L)$, though present, are drastically
reduced for $3d$, in contrast to those in $1d$ and $2d$
[Figs.\ref{fig:Conductance}(a),(c) and Fig.\ref{fig:Conductance_HigherD}(a)]. A
critical point at $V=V_c\simeq 2.2$ can be clearly detected from the crossing
of curves as function of $V$ for different system sizes, as shown in the inset
of Fig.\ref{fig:Conductance_HigherD}(c). The crossing also indicates a scale
invariant conductance at the critical point. A reasonably good scaling collapse
of the data using a single-parameter finite-size scaling form
$\ln{(g_\mr{K}(L))}=\mathcal{F}((V_c-V)L^{1/\nu})$ could be obtained near the
critical point, as shown in Fig.\ref{fig:Conductance_HigherD}(c). The
finite-size scaling yields $\nu=1.60\pm 0.04$ and $V_c=2.22\pm 0.01$,
consistent with earlier study in ref.\cite{Devakul2017} using multifractal
finite-size scaling analysis of wave function of closed system. The universal
scaling curve describes the $g_\mr{K}(L,V)$ data quite well as shown by the
solid lines in Figs.\ref{fig:Conductance_HigherD}(b) and (c)(inset). This is in
tune with a continuous metal-insulator transition in $3d$, unlike those in the
$1d$ and $2d$ quasiperiodic systems. Moreover, this is consistent with
single-parameter scaling law, $\beta(g)=d\ln g/d\ln L$, in the weaker sense
\cite{Slevin2001} in the $3d$ quasiperiodic system. However, the persistence of
weak non-monotonic system-size variations in the typical conductance still
violates the assumption of monotonicity of $\beta(g)$ in the scaling theory
\cite{EAbrahams1979}. The weak non monotonicity, though, could be due to
limited system sizes accessed in $3d$ and one might recover strict
single-parameter scaling at larger lengths. 

From the real space-momentum space duality of the model [\eqref{eq:model}], we
expect another transition around $V\sim 1/V_c\approx 0.45$ from a diffusive to
a ballistic phase \cite{Devakul2017}. Our results do not show any transport
signature of this transition. As shown in Fig.\ref{fig:Conductance_HigherD}(d),
$g_\mr{K}(L)$ around $V= 0.45$ can be well described by superdiffusive length
scaling with an exponent $\alpha>1$. This could be due to the fact that the
duality is not strictly valid for such finite system connected to leads and due
to the dichotomy between open and closed system properties, as seen in the $1d$
quasiperiodic system \cite{Purkayastha2017,Purkayastha2018}.

\subsection{Beta function} 

As already remarked, the strong non-monotonicity of even typical $g(L)$ in $1d$
and $2d$ invalidates the application of single-parameter scaling. However, we
construct a $\beta(\tilde{g})$, where $\tilde{g}(L)$ is extracted from fitting
a smooth curve to the data for $g_\mr{T}(L)$ and $g_\mr{L}(L)$, e.g. the ones
shown in Fig.\ref{fig:Conductance}, for several values of $V$ shown in
Figs.\ref{fig:BetaFunction}(a),(b). This is rather unambiguous procedure in
$1d$ where the overall $L$ dependences of conductance in the localized,
critical and metallic states are very well described by exponentially
localized, subdiffusive and ballistic behaviors, respectively, over several
decades of $L$ [Figs.\ref{fig:Conductance}(a),(c)]. The results for the
respective beta functions $\beta(\tilde{g})$ in $1d$ are shown in
Figs.\ref{fig:BetaFunction}(a),(b) across the metal-insulator transition. In
Fig.\ref{fig:BetaFunction}(a), $\beta(\tilde{g}_\mr{T})$ has separate curves
for individual phases and the critical point as well as for different
sequences.  For example, the multiple straight lines at the critical value
$V=1$ are due to distinct powerlaws for different sequences shown in
Fig.\ref{fig:Conductance}(b). These, and the jump of beta functions across the
critical point clearly violates the assumption of continuity in the
single-parameter scaling theory. Similar features are seen in
$\beta(\tilde{g}_\mr{L})$ [Fig.\ref{fig:BetaFunction}(b)]. We find
$\tilde{g}_\mr{L}(V,L)=g_0(V) \exp(-L/\xi(V))$ to describe quite accurately the
conductance in the localized phase, even very close to the transition. However,
the coefficient $g_0$, a measure of conductance at the microscopic scale
$\ell$, varies substantially with $V$ [see inset of
Fig.\ref{fig:BetaFunction}(b)]. This is unlike, e.g., that in the $1d$ Anderson
model where $g_0\approx 1$ irrespective of the disorder strength. As a result,
one can only obtain a universal $\beta(\tilde{g})$ curve for the localized
phase in $1d$  as a function of $\ln(\tilde{g}_\mr{L}/g_0(V))$, i.e. after
dividing $g_\mr{L}$ with appropriate $g_0$. 

To contrast the above results for beta function for $1d$ quasiperiodic system
with that of random system, we show in the Supplementary Information that even
a small amount of randomness, introduced, e.g., by elevating the phase $\phi$
to a random variable at each site,  makes $g_\mr{L}(L)$  exponentially decaying
with $L$ but with small non-monotonicity and hence leads to a continuous beta
function for the overall conductance.

As shown in Fig.\ref{fig:BetaFunction}(c), we find very similar result for
$\beta(\tilde{g})$ in $2d$, extracted from, e.g., the fitting curves in
Fig.\ref{fig:Conductance_HigherD}(a). Here the beta function also jumps from a
localized behaviour, $\beta(\tilde{g})\propto \ln(\tilde{g}/g_0)$, to a
constant superdiffusive value $\beta(\tilde{g})\simeq 0.3$ in the metallic
phase, across a subdiffusive critical state with $\beta(\tilde{g})\simeq-0.52$.
However, as commented earlier, the asymptotic scaling behaviors in $2d$ can
only be extracted for $L$ above a substantially large microscopic length scale
$\ell$ and hence the beta functions are extracted from only a limited ranges of
system sizes. 

Both the $1d$ and $2d$ results indicate strong violation of the assumption of
continuity of $\beta(g)$ in single-parameter scaling theory, even when we
disregard the non monotonicity of $g(L)$ by extracting an overall smooth
$\tilde{g}(L)$ from the asymptotic behaviors at large system sizes $L$. The
above procedure can not be carried out in $3d$ close to the critical point,
since our system sizes are limited to much smaller values of $L\leq 30$.
However, since the non-monotonicity of $g(L)$ is much weaker in $3d$ and a
reasonable scaling collapse of the data could be obtained near the
metal-insulator transition, we extract the $\beta(g)$ in
Fig.\ref{fig:BetaFunction}(d) (dashed black line) near the transition from the
scaling fit of $g_\mr{K}(L)$, shown in
Figs.\ref{fig:Conductance_HigherD}(b),(c). The fit describes the data well over
reasonably large range of $V$ and $L$ and hence suggests the restoration of
continuity of $\beta(g)$ for the $3d$ quasiperiodic system, provided we neglect
the weak non-monotonic variations of $g(L)$. In Fig.\ref{fig:BetaFunction}(d),
we also show that the beta function extracted from exponential fit deep in the
insulating phase and from powerlaw fits deep in the metallic phase is
consistent with that obtained from scaling collapse near the transition.

\section{Conclusions} \label{sec:conclusion}

In summary, we have studied transport properties in a particular class of
self-dual quasiperiodic models in one, two, an three dimensions. We have
focussed on the system size dependences of the Thouless and open-system
Landauer/Kubo conductances.  Our results uncover the intricate nature of
transport in quasiperiodic systems, which is manifested in terms of the
non-monotonic system-size dependence of typical conductances, e.g., because of
transport resonances, and a variety of sub-diffusive power laws for critical
transport; these depend on the dimension, energy, and the sequences of length
we have described above. 

Our results reveal the absence of a single-parameter-scaling description in low
dimensions and a recovery of weak single-parameter scaling in $3d$; this has
direct implications for universality classes of metal-insulator transition in
quasiperiodic systems. We plan to  compute the multifractal spectrum of the
wavefunction and the Thouless conductance at the critical point in the $3d$
quasiperiodic model and compare it with those at the $3d$ Anderson transition
to verify whether they truly belong to the same universality class. It would
also be worthwhile to look into generalizations of quasiperiodic systems to
other symmetry classes~\cite{Devakul2017} from this perspective. Morover, it
would also be interesting to study the implications of sub-diffusive critical
states of the non-interacting models, specially in $1d$, on the Griffith-like
effect seen experimentally near the MBL transition in interacting quasiperiodic
system~\cite{Luschen2017} and incorporate these critical states into a
real-space-renormalization-group framework \cite{Vosk2015,Potter2015,Zhang2018}
for the MBL transition in quasiperiodic systems.   

\appendix
\section{Higher-dimensional generalization of Aubry-Andre model} \label{app:HigherDModel}
We study the model proposed in Ref.~\cite{Devakul2017} as a generalization of 
the self-dual $1d$ Aubry-Andre model to $d$ dimensions, namely,
\begin{subequations}\label{eq:model}
\begin{align}
\mathcal{H}&=t\sum_{\mathbf{r},\mu}\left(e^{i\phi_\mu}c^\dagger_{\mathbf{r}+\hat{\mathbf{\mu}}}c_\mathbf{r}+\mathrm{h.c}\right)+\sum_\mathbf{r}\epsilon_\mathbf{r}c^\dagger_\mathbf{r}c_\mathbf{r} \label{eq:modelHamiltonian}\\
\epsilon_\mathbf{r}&=2V\sum_{\mu=1}^{d}\cos\left(2\pi\sum_{\nu=1}^d B_{\mu\nu}r_\nu+\phi_\mu\right) \label{eq:potential}
\end{align}
\end{subequations}
Where $c_\mathbf{r}$ is the fermion operator at site $\mathbf{r}$ of a $d$-dimensional hypercubic lattice and $\mu=1,\dots,d$ denotes Cartesian components. We choose $t=1$, the matrix $\mathbb{B}=b\mathbb{R}$ with $b=\Phi$ and an orthonormal matrix $\mathbb{R}$ \cite{Devakul2017}. In $1d$, $\mathbb{R}=1$ and
\begin{subequations}
\begin{align}
\mathbb{R}&=\begin{bmatrix}
c & -s\\
s & c \end{bmatrix} ~~~~~~d=2\\
\mathbb{R}&=\begin{bmatrix}
c^2+s^3 & cs & cs^2-cs\\
cs & -s & c^2 \\
cs^2-cs & c^2 & c^2s+s^2 
\end{bmatrix}~~~~~~d=3
\end{align}
\end{subequations}
where $c=\cos\theta$ and $s=\sin\theta$. We choose $\theta=\pi/7$ for all our calculations. For the calculations of conductance of open system connected with leads, we use free boundary condition in transverse directions and hence the phase factor in the hopping term of Eq.\eqref{eq:model} can be gauged away. To compare the open system conductance with that of the closed one, we consider the Hamiltonian again without the phase factor in the hopping to calculate the Thouless conductance.  We note that for the above transport set up for a finite system the real-momentum space duality of the model Eq.[\eqref{eq:model}] \cite {Devakul2017} is lost. For each finite system with linear dimension $L$ under periodic boundary condition one can generate a self-dual approximation \cite{Devakul2017}. This recipe, however, is not applicable for the transport set up. All the data points for the quasi periodic system, shown here and in the \textit{Supplementary Information}, are results of averaging over $300-400$ points of $\phi\in [0,2\pi)$ and we checked the convergence of these data for several parameter values with larger number ($\sim 1000-2000$) of $\phi$ averages.


%

\subsection*{Acknowledgement} We thank Sriram Ganeshan, Shivaji Sondhi, Archak
Purakayastha and Chandan Dasgupta for useful discussions. SB acknowledges
support from The Infosys Foundation, India. SM acknowledges support from the
Indo-Israeli ISF-UGC grant. RP acknowledges support from DST (India).

\bibliography{Localization}

\begin{thebibliography}{50}%
\makeatletter
\providecommand \@ifxundefined [1]{%
 \@ifx{#1\undefined}
}%
\providecommand \@ifnum [1]{%
 \ifnum #1\expandafter \@firstoftwo
 \else \expandafter \@secondoftwo
 \fi
}%
\providecommand \@ifx [1]{%
 \ifx #1\expandafter \@firstoftwo
 \else \expandafter \@secondoftwo
 \fi
}%
\providecommand \natexlab [1]{#1}%
\providecommand \enquote  [1]{``#1''}%
\providecommand \bibnamefont  [1]{#1}%
\providecommand \bibfnamefont [1]{#1}%
\providecommand \citenamefont [1]{#1}%
\providecommand \href@noop [0]{\@secondoftwo}%
\providecommand \href [0]{\begingroup \@sanitize@url \@href}%
\providecommand \@href[1]{\@@startlink{#1}\@@href}%
\providecommand \@@href[1]{\endgroup#1\@@endlink}%
\providecommand \@sanitize@url [0]{\catcode `\\12\catcode `\$12\catcode
  `\&12\catcode `\#12\catcode `\^12\catcode `\_12\catcode `\%12\relax}%
\providecommand \@@startlink[1]{}%
\providecommand \@@endlink[0]{}%
\providecommand \url  [0]{\begingroup\@sanitize@url \@url }%
\providecommand \@url [1]{\endgroup\@href {#1}{\urlprefix }}%
\providecommand \urlprefix  [0]{URL }%
\providecommand \Eprint [0]{\href }%
\providecommand \doibase [0]{http://dx.doi.org/}%
\providecommand \selectlanguage [0]{\@gobble}%
\providecommand \bibinfo  [0]{\@secondoftwo}%
\providecommand \bibfield  [0]{\@secondoftwo}%
\providecommand \translation [1]{[#1]}%
\providecommand \BibitemOpen [0]{}%
\providecommand \bibitemStop [0]{}%
\providecommand \bibitemNoStop [0]{.\EOS\space}%
\providecommand \EOS [0]{\spacefactor3000\relax}%
\providecommand \BibitemShut  [1]{\csname bibitem#1\endcsname}%
\let\auto@bib@innerbib\@empty
\bibitem [{\citenamefont {Abrahams}\ \emph {et~al.}(1979)\citenamefont
  {Abrahams}, \citenamefont {Anderson}, \citenamefont {Licciardello},\ and\
  \citenamefont {Ramakrishnan}}]{EAbrahams1979}%
  \BibitemOpen
  \bibfield  {author} {\bibinfo {author} {\bibfnamefont {E.}~\bibnamefont
  {Abrahams}}, \bibinfo {author} {\bibfnamefont {P.~W.}\ \bibnamefont
  {Anderson}}, \bibinfo {author} {\bibfnamefont {D.~C.}\ \bibnamefont
  {Licciardello}}, \ and\ \bibinfo {author} {\bibfnamefont {T.~V.}\
  \bibnamefont {Ramakrishnan}},\ }\bibfield  {title} {\enquote {\bibinfo
  {title} {Scaling theory of localization: Absence of quantum diffusion in two
  dimensions},}\ }\href {\doibase 10.1103/PhysRevLett.42.673} {\bibfield
  {journal} {\bibinfo  {journal} {Phys. Rev. Lett.}\ }\textbf {\bibinfo
  {volume} {42}},\ \bibinfo {pages} {673--676} (\bibinfo {year}
  {1979})}\BibitemShut {NoStop}%
\bibitem [{\citenamefont {Anderson}(1958)}]{PWAnderson1958}%
  \BibitemOpen
  \bibfield  {author} {\bibinfo {author} {\bibfnamefont {P.~W.}\ \bibnamefont
  {Anderson}},\ }\bibfield  {title} {\enquote {\bibinfo {title} {Absence of
  diffusion in certain random lattices},}\ }\href {\doibase
  10.1103/PhysRev.109.1492} {\bibfield  {journal} {\bibinfo  {journal} {Phys.
  Rev.}\ }\textbf {\bibinfo {volume} {109}},\ \bibinfo {pages} {1492--1505}
  (\bibinfo {year} {1958})}\BibitemShut {NoStop}%
\bibitem [{\citenamefont {Aubry}\ and\ \citenamefont
  {Andre}(1980)}]{Aubry1980}%
  \BibitemOpen
  \bibfield  {author} {\bibinfo {author} {\bibfnamefont {S.}~\bibnamefont
  {Aubry}}\ and\ \bibinfo {author} {\bibfnamefont {G.}~\bibnamefont {Andre}},\
  }\bibfield  {title} {\enquote {\bibinfo {title} {Analyticity breaking and
  anderson localization in incommensurate lattices},}\ }\href@noop {}
  {\bibfield  {journal} {\bibinfo  {journal} {Ann. Israel Phys. Soc.}\ }\textbf
  {\bibinfo {volume} {3}},\ \bibinfo {pages} {18} (\bibinfo {year}
  {1980})}\BibitemShut {NoStop}%
\bibitem [{\citenamefont {Simon}(1982)}]{Simon1982}%
  \BibitemOpen
  \bibfield  {author} {\bibinfo {author} {\bibfnamefont {Barry}\ \bibnamefont
  {Simon}},\ }\bibfield  {title} {\enquote {\bibinfo {title} {{Almost periodic
  Schrödinger operators: A Review}},}\ }\href {\doibase
  https://doi.org/10.1016/S0196-8858(82)80018-3} {\bibfield  {journal}
  {\bibinfo  {journal} {Advances in Applied Mathematics}\ }\textbf {\bibinfo
  {volume} {3}},\ \bibinfo {pages} {463 -- 490} (\bibinfo {year}
  {1982})}\BibitemShut {NoStop}%
\bibitem [{\citenamefont {Sokoloff}\ and\ \citenamefont
  {Jos\'e}(1982)}]{Sokoloff1982}%
  \BibitemOpen
  \bibfield  {author} {\bibinfo {author} {\bibfnamefont {J.~B.}\ \bibnamefont
  {Sokoloff}}\ and\ \bibinfo {author} {\bibfnamefont {Jorge~V.}\ \bibnamefont
  {Jos\'e}},\ }\bibfield  {title} {\enquote {\bibinfo {title} {Localization in
  an almost periodically modulated array of potential barriers},}\ }\href
  {\doibase 10.1103/PhysRevLett.49.334} {\bibfield  {journal} {\bibinfo
  {journal} {Phys. Rev. Lett.}\ }\textbf {\bibinfo {volume} {49}},\ \bibinfo
  {pages} {334--337} (\bibinfo {year} {1982})}\BibitemShut {NoStop}%
\bibitem [{\citenamefont {Thouless}\ and\ \citenamefont
  {Niu}(1983)}]{Thouless1983}%
  \BibitemOpen
  \bibfield  {author} {\bibinfo {author} {\bibfnamefont {D.~J.}\ \bibnamefont
  {Thouless}}\ and\ \bibinfo {author} {\bibfnamefont {Q.}~\bibnamefont {Niu}},\
  }\bibfield  {title} {\enquote {\bibinfo {title} {Wavefunction scaling in a
  quasi-periodic potential},}\ }\href
  {http://stacks.iop.org/0305-4470/16/i=9/a=015} {\bibfield  {journal}
  {\bibinfo  {journal} {Journal of Physics A: Mathematical and General}\
  }\textbf {\bibinfo {volume} {16}},\ \bibinfo {pages} {1911} (\bibinfo {year}
  {1983})}\BibitemShut {NoStop}%
\bibitem [{\citenamefont {Kohmoto}\ \emph {et~al.}(1983)\citenamefont
  {Kohmoto}, \citenamefont {Kadanoff},\ and\ \citenamefont
  {Tang}}]{Kohmoto1983}%
  \BibitemOpen
  \bibfield  {author} {\bibinfo {author} {\bibfnamefont {M.}~\bibnamefont
  {Kohmoto}}, \bibinfo {author} {\bibfnamefont {L.~P.}\ \bibnamefont
  {Kadanoff}}, \ and\ \bibinfo {author} {\bibfnamefont {C.}~\bibnamefont
  {Tang}},\ }\bibfield  {title} {\enquote {\bibinfo {title} {Localization
  problem in one dimension: Mapping and escape},}\ }\href {\doibase
  10.1103/PhysRevLett.50.1870} {\bibfield  {journal} {\bibinfo  {journal}
  {Phys. Rev. Lett.}\ }\textbf {\bibinfo {volume} {50}},\ \bibinfo {pages}
  {1870--1872} (\bibinfo {year} {1983})}\BibitemShut {NoStop}%
\bibitem [{\citenamefont {Kohmoto}(1983)}]{Kohmoto1983_1}%
  \BibitemOpen
  \bibfield  {author} {\bibinfo {author} {\bibfnamefont {M.}~\bibnamefont
  {Kohmoto}},\ }\bibfield  {title} {\enquote {\bibinfo {title} {Metal-insulator
  transition and scaling for incommensurate systems},}\ }\href {\doibase
  10.1103/PhysRevLett.51.1198} {\bibfield  {journal} {\bibinfo  {journal}
  {Phys. Rev. Lett.}\ }\textbf {\bibinfo {volume} {51}},\ \bibinfo {pages}
  {1198--1201} (\bibinfo {year} {1983})}\BibitemShut {NoStop}%
\bibitem [{\citenamefont {Ostlund}\ \emph {et~al.}(1983)\citenamefont
  {Ostlund}, \citenamefont {Pandit}, \citenamefont {Rand}, \citenamefont
  {Schellnhuber},\ and\ \citenamefont {Siggia}}]{Ostlund1983}%
  \BibitemOpen
  \bibfield  {author} {\bibinfo {author} {\bibfnamefont {S.}~\bibnamefont
  {Ostlund}}, \bibinfo {author} {\bibfnamefont {R.}~\bibnamefont {Pandit}},
  \bibinfo {author} {\bibfnamefont {D.}~\bibnamefont {Rand}}, \bibinfo {author}
  {\bibfnamefont {H.~J.}\ \bibnamefont {Schellnhuber}}, \ and\ \bibinfo
  {author} {\bibfnamefont {E.~D.}\ \bibnamefont {Siggia}},\ }\bibfield  {title}
  {\enquote {\bibinfo {title} {{One-Dimensional Schr\"odinger Equation with an
  Almost Periodic Potential}},}\ }\href {\doibase 10.1103/PhysRevLett.50.1873}
  {\bibfield  {journal} {\bibinfo  {journal} {Phys. Rev. Lett.}\ }\textbf
  {\bibinfo {volume} {50}},\ \bibinfo {pages} {1873--1876} (\bibinfo {year}
  {1983})}\BibitemShut {NoStop}%
\bibitem [{\citenamefont {Ostlund}\ and\ \citenamefont
  {Pandit}(1984)}]{Ostlund1984}%
  \BibitemOpen
  \bibfield  {author} {\bibinfo {author} {\bibfnamefont {S.}~\bibnamefont
  {Ostlund}}\ and\ \bibinfo {author} {\bibfnamefont {R.}~\bibnamefont
  {Pandit}},\ }\bibfield  {title} {\enquote {\bibinfo {title}
  {{Renormalization-group analysis of the discrete quasiperiodic Schr\"odinger
  equation}},}\ }\href {\doibase 10.1103/PhysRevB.29.1394} {\bibfield
  {journal} {\bibinfo  {journal} {Phys. Rev. B}\ }\textbf {\bibinfo {volume}
  {29}},\ \bibinfo {pages} {1394--1414} (\bibinfo {year} {1984})}\BibitemShut
  {NoStop}%
\bibitem [{\citenamefont {Sokoloff}(1985)}]{Sokoloff1985}%
  \BibitemOpen
  \bibfield  {author} {\bibinfo {author} {\bibfnamefont {J.~B.\textbf{•}}\
  \bibnamefont {Sokoloff}},\ }\bibfield  {title} {\enquote {\bibinfo {title}
  {Unusual band structure, wave functions and electrical conductance in
  crystals with incommensurate periodic potentials},}\ }\href {\doibase
  https://doi.org/10.1016/0370-1573(85)90088-2} {\bibfield  {journal} {\bibinfo
   {journal} {Physics Reports}\ }\textbf {\bibinfo {volume} {126}},\ \bibinfo
  {pages} {189 -- 244} (\bibinfo {year} {1985})}\BibitemShut {NoStop}%
\bibitem [{\citenamefont {{Schreiber}}\ \emph {et~al.}(2015)\citenamefont
  {{Schreiber}}, \citenamefont {{Hodgman}}, \citenamefont {{Bordia}},
  \citenamefont {{L{\"u}schen}}, \citenamefont {{Fischer}}, \citenamefont
  {{Vosk}}, \citenamefont {{Altman}}, \citenamefont {{Schneider}},\ and\
  \citenamefont {{Bloch}}}]{Schreiber2015}%
  \BibitemOpen
  \bibfield  {author} {\bibinfo {author} {\bibfnamefont {M.}~\bibnamefont
  {{Schreiber}}}, \bibinfo {author} {\bibfnamefont {S.~S.}\ \bibnamefont
  {{Hodgman}}}, \bibinfo {author} {\bibfnamefont {P.}~\bibnamefont {{Bordia}}},
  \bibinfo {author} {\bibfnamefont {H.~P.}\ \bibnamefont {{L{\"u}schen}}},
  \bibinfo {author} {\bibfnamefont {M.~H.}\ \bibnamefont {{Fischer}}}, \bibinfo
  {author} {\bibfnamefont {R.}~\bibnamefont {{Vosk}}}, \bibinfo {author}
  {\bibfnamefont {E.}~\bibnamefont {{Altman}}}, \bibinfo {author}
  {\bibfnamefont {U.}~\bibnamefont {{Schneider}}}, \ and\ \bibinfo {author}
  {\bibfnamefont {I.}~\bibnamefont {{Bloch}}},\ }\bibfield  {title} {\enquote
  {\bibinfo {title} {{Observation of many-body localization of interacting
  fermions in a quasirandom optical lattice}},}\ }\href {\doibase
  10.1126/science.aaa7432} {\bibfield  {journal} {\bibinfo  {journal}
  {Science}\ }\textbf {\bibinfo {volume} {349}},\ \bibinfo {pages} {842--845}
  (\bibinfo {year} {2015})}\BibitemShut {NoStop}%
\bibitem [{\citenamefont {Purkayastha}\ \emph {et~al.}(2017)\citenamefont
  {Purkayastha}, \citenamefont {Dhar},\ and\ \citenamefont
  {Kulkarni}}]{Purkayastha2017}%
  \BibitemOpen
  \bibfield  {author} {\bibinfo {author} {\bibfnamefont {A.}~\bibnamefont
  {Purkayastha}}, \bibinfo {author} {\bibfnamefont {A.}~\bibnamefont {Dhar}}, \
  and\ \bibinfo {author} {\bibfnamefont {M.}~\bibnamefont {Kulkarni}},\
  }\bibfield  {title} {\enquote {\bibinfo {title} {Nonequilibrium phase diagram
  of a one-dimensional quasiperiodic system with a single-particle mobility
  edge},}\ }\href {\doibase 10.1103/PhysRevB.96.180204} {\bibfield  {journal}
  {\bibinfo  {journal} {Phys. Rev. B}\ }\textbf {\bibinfo {volume} {96}},\
  \bibinfo {pages} {180204} (\bibinfo {year} {2017})}\BibitemShut {NoStop}%
\bibitem [{\citenamefont {Purkayastha}\ \emph {et~al.}(2018)\citenamefont
  {Purkayastha}, \citenamefont {Sanyal}, \citenamefont {Dhar},\ and\
  \citenamefont {Kulkarni}}]{Purkayastha2018}%
  \BibitemOpen
  \bibfield  {author} {\bibinfo {author} {\bibfnamefont {A.}~\bibnamefont
  {Purkayastha}}, \bibinfo {author} {\bibfnamefont {S.}~\bibnamefont {Sanyal}},
  \bibinfo {author} {\bibfnamefont {A.}~\bibnamefont {Dhar}}, \ and\ \bibinfo
  {author} {\bibfnamefont {M.}~\bibnamefont {Kulkarni}},\ }\bibfield  {title}
  {\enquote {\bibinfo {title} {{Anomalous transport in the Aubry-Andr\'e-Harper
  model in isolated and open systems}},}\ }\href {\doibase
  10.1103/PhysRevB.97.174206} {\bibfield  {journal} {\bibinfo  {journal} {Phys.
  Rev. B}\ }\textbf {\bibinfo {volume} {97}},\ \bibinfo {pages} {174206}
  (\bibinfo {year} {2018})}\BibitemShut {NoStop}%
\bibitem [{\citenamefont {Varma}\ \emph {et~al.}(2017)\citenamefont {Varma},
  \citenamefont {de~Mulatier},\ and\ \citenamefont {\ifmmode \check{Z}\else
  \v{Z}\fi{}nidari\ifmmode~\check{c}\else \v{c}\fi{}}}]{Varma2017}%
  \BibitemOpen
  \bibfield  {author} {\bibinfo {author} {\bibfnamefont {V.~K.}\ \bibnamefont
  {Varma}}, \bibinfo {author} {\bibfnamefont {C.}~\bibnamefont {de~Mulatier}},
  \ and\ \bibinfo {author} {\bibfnamefont {M.}~\bibnamefont {\ifmmode
  \check{Z}\else \v{Z}\fi{}nidari\ifmmode~\check{c}\else \v{c}\fi{}}},\
  }\bibfield  {title} {\enquote {\bibinfo {title} {Fractality in nonequilibrium
  steady states of quasiperiodic systems},}\ }\href {\doibase
  10.1103/PhysRevE.96.032130} {\bibfield  {journal} {\bibinfo  {journal} {Phys.
  Rev. E}\ }\textbf {\bibinfo {volume} {96}},\ \bibinfo {pages} {032130}
  (\bibinfo {year} {2017})}\BibitemShut {NoStop}%
\bibitem [{\citenamefont {Khemani}\ \emph {et~al.}(2017)\citenamefont
  {Khemani}, \citenamefont {Sheng},\ and\ \citenamefont {Huse}}]{Khemani2017}%
  \BibitemOpen
  \bibfield  {author} {\bibinfo {author} {\bibfnamefont {V.}~\bibnamefont
  {Khemani}}, \bibinfo {author} {\bibfnamefont {D.~N.}\ \bibnamefont {Sheng}},
  \ and\ \bibinfo {author} {\bibfnamefont {D.~A.}\ \bibnamefont {Huse}},\
  }\bibfield  {title} {\enquote {\bibinfo {title} {Two universality classes for
  the many-body localization transition},}\ }\href {\doibase
  10.1103/PhysRevLett.119.075702} {\bibfield  {journal} {\bibinfo  {journal}
  {Phys. Rev. Lett.}\ }\textbf {\bibinfo {volume} {119}},\ \bibinfo {pages}
  {075702} (\bibinfo {year} {2017})}\BibitemShut {NoStop}%
\bibitem [{\citenamefont {De~Roeck}\ and\ \citenamefont
  {Huveneers}(2017)}]{DeRoeck2016}%
  \BibitemOpen
  \bibfield  {author} {\bibinfo {author} {\bibfnamefont {Wojciech}\
  \bibnamefont {De~Roeck}}\ and\ \bibinfo {author} {\bibfnamefont {Fran\ifmmode
  \mbox{\c{c}}\else~\c{c}\fi{}ois}\ \bibnamefont {Huveneers}},\ }\bibfield
  {title} {\enquote {\bibinfo {title} {Stability and instability towards
  delocalization in many-body localization systems},}\ }\href {\doibase
  10.1103/PhysRevB.95.155129} {\bibfield  {journal} {\bibinfo  {journal} {Phys.
  Rev. B}\ }\textbf {\bibinfo {volume} {95}},\ \bibinfo {pages} {155129}
  (\bibinfo {year} {2017})}\BibitemShut {NoStop}%
\bibitem [{\citenamefont {{Potirniche}}\ \emph {et~al.}(2018)\citenamefont
  {{Potirniche}}, \citenamefont {{Banerjee}},\ and\ \citenamefont
  {{Altman}}}]{Potriniche2018}%
  \BibitemOpen
  \bibfield  {author} {\bibinfo {author} {\bibfnamefont {I.-D.}\ \bibnamefont
  {{Potirniche}}}, \bibinfo {author} {\bibfnamefont {S.}~\bibnamefont
  {{Banerjee}}}, \ and\ \bibinfo {author} {\bibfnamefont {E.}~\bibnamefont
  {{Altman}}},\ }\bibfield  {title} {\enquote {\bibinfo {title} {{On the
  stability of many-body localization in $d>1$}},}\ }\href@noop {} {\bibfield
  {journal} {\bibinfo  {journal} {ArXiv e-prints}\ } (\bibinfo {year}
  {2018})},\ \Eprint {http://arxiv.org/abs/1805.01475} {arXiv:1805.01475
  [cond-mat.dis-nn]} \BibitemShut {NoStop}%
\bibitem [{\citenamefont {Evers}\ and\ \citenamefont
  {Mirlin}(2008)}]{Evers2008}%
  \BibitemOpen
  \bibfield  {author} {\bibinfo {author} {\bibfnamefont {F.}~\bibnamefont
  {Evers}}\ and\ \bibinfo {author} {\bibfnamefont {A.~D.}\ \bibnamefont
  {Mirlin}},\ }\bibfield  {title} {\enquote {\bibinfo {title} {Anderson
  transitions},}\ }\href {\doibase 10.1103/RevModPhys.80.1355} {\bibfield
  {journal} {\bibinfo  {journal} {Rev. Mod. Phys.}\ }\textbf {\bibinfo {volume}
  {80}},\ \bibinfo {pages} {1355--1417} (\bibinfo {year} {2008})}\BibitemShut
  {NoStop}%
\bibitem [{\citenamefont {Anderson}\ \emph {et~al.}(1980)\citenamefont
  {Anderson}, \citenamefont {Thouless}, \citenamefont {Abrahams},\ and\
  \citenamefont {Fisher}}]{Anderson1980}%
  \BibitemOpen
  \bibfield  {author} {\bibinfo {author} {\bibfnamefont {P.~W.}\ \bibnamefont
  {Anderson}}, \bibinfo {author} {\bibfnamefont {D.~J.}\ \bibnamefont
  {Thouless}}, \bibinfo {author} {\bibfnamefont {E.}~\bibnamefont {Abrahams}},
  \ and\ \bibinfo {author} {\bibfnamefont {D.~S.}\ \bibnamefont {Fisher}},\
  }\bibfield  {title} {\enquote {\bibinfo {title} {New method for a scaling
  theory of localization},}\ }\href {\doibase 10.1103/PhysRevB.22.3519}
  {\bibfield  {journal} {\bibinfo  {journal} {Phys. Rev. B}\ }\textbf {\bibinfo
  {volume} {22}},\ \bibinfo {pages} {3519--3526} (\bibinfo {year}
  {1980})}\BibitemShut {NoStop}%
\bibitem [{\citenamefont {Altshuler}(1985)}]{Altshuler1985}%
  \BibitemOpen
  \bibfield  {author} {\bibinfo {author} {\bibfnamefont {B.~L.}\ \bibnamefont
  {Altshuler}},\ }\bibfield  {title} {\enquote {\bibinfo {title} {Fluctuations
  in the extrinsic conductivity of disordered conductors},}\ }\href@noop {}
  {\bibfield  {journal} {\bibinfo  {journal} {JETP Lett.}\ }\textbf {\bibinfo
  {volume} {41}},\ \bibinfo {pages} {648} (\bibinfo {year} {1985})}\BibitemShut
  {NoStop}%
\bibitem [{\citenamefont {Lee}\ and\ \citenamefont {Stone}(1985)}]{Lee1985}%
  \BibitemOpen
  \bibfield  {author} {\bibinfo {author} {\bibfnamefont {P.~A.}\ \bibnamefont
  {Lee}}\ and\ \bibinfo {author} {\bibfnamefont {A.~D.}\ \bibnamefont
  {Stone}},\ }\bibfield  {title} {\enquote {\bibinfo {title} {Universal
  conductance fluctuations in metals},}\ }\href {\doibase
  10.1103/PhysRevLett.55.1622} {\bibfield  {journal} {\bibinfo  {journal}
  {Phys. Rev. Lett.}\ }\textbf {\bibinfo {volume} {55}},\ \bibinfo {pages}
  {1622--1625} (\bibinfo {year} {1985})}\BibitemShut {NoStop}%
\bibitem [{\citenamefont {Lee}\ and\ \citenamefont {Fisher}(1981)}]{Lee1981}%
  \BibitemOpen
  \bibfield  {author} {\bibinfo {author} {\bibfnamefont {P.~A.}\ \bibnamefont
  {Lee}}\ and\ \bibinfo {author} {\bibfnamefont {D.~S.}\ \bibnamefont
  {Fisher}},\ }\bibfield  {title} {\enquote {\bibinfo {title} {Anderson
  localization in two dimensions},}\ }\href {\doibase
  10.1103/PhysRevLett.47.882} {\bibfield  {journal} {\bibinfo  {journal} {Phys.
  Rev. Lett.}\ }\textbf {\bibinfo {volume} {47}},\ \bibinfo {pages} {882--885}
  (\bibinfo {year} {1981})}\BibitemShut {NoStop}%
\bibitem [{\citenamefont {Pichard}\ and\ \citenamefont
  {Sarma}(1981)}]{Pichard1981}%
  \BibitemOpen
  \bibfield  {author} {\bibinfo {author} {\bibfnamefont {J.~L.}\ \bibnamefont
  {Pichard}}\ and\ \bibinfo {author} {\bibfnamefont {G.}~\bibnamefont
  {Sarma}},\ }\bibfield  {title} {\enquote {\bibinfo {title} {Finite-size
  scaling approach to anderson localisation. ii. quantitative analysis and new
  results},}\ }\href {http://stacks.iop.org/0022-3719/14/i=21/a=004} {\bibfield
   {journal} {\bibinfo  {journal} {Journal of Physics C: Solid State Physics}\
  }\textbf {\bibinfo {volume} {14}},\ \bibinfo {pages} {L617} (\bibinfo {year}
  {1981})}\BibitemShut {NoStop}%
\bibitem [{\citenamefont {MacKinnon}\ and\ \citenamefont
  {Kramer}(1983)}]{MacKinnon1983}%
  \BibitemOpen
  \bibfield  {author} {\bibinfo {author} {\bibfnamefont {A.}~\bibnamefont
  {MacKinnon}}\ and\ \bibinfo {author} {\bibfnamefont {B.}~\bibnamefont
  {Kramer}},\ }\bibfield  {title} {\enquote {\bibinfo {title} {The scaling
  theory of electrons in disordered solids: Additional numerical results},}\
  }\href {\doibase 10.1007/BF01578242} {\bibfield  {journal} {\bibinfo
  {journal} {Zeitschrift f{\"u}r Physik B Condensed Matter}\ }\textbf {\bibinfo
  {volume} {53}},\ \bibinfo {pages} {1--13} (\bibinfo {year}
  {1983})}\BibitemShut {NoStop}%
\bibitem [{\citenamefont {Slevin}\ \emph {et~al.}(2001)\citenamefont {Slevin},
  \citenamefont {Marko\ifmmode~\check{s}\else \v{s}\fi{}},\ and\ \citenamefont
  {Ohtsuki}}]{Slevin2001}%
  \BibitemOpen
  \bibfield  {author} {\bibinfo {author} {\bibfnamefont {K.}~\bibnamefont
  {Slevin}}, \bibinfo {author} {\bibfnamefont {P.}~\bibnamefont
  {Marko\ifmmode~\check{s}\else \v{s}\fi{}}}, \ and\ \bibinfo {author}
  {\bibfnamefont {T.}~\bibnamefont {Ohtsuki}},\ }\bibfield  {title} {\enquote
  {\bibinfo {title} {Reconciling conductance fluctuations and the scaling
  theory of localization},}\ }\href {\doibase 10.1103/PhysRevLett.86.3594}
  {\bibfield  {journal} {\bibinfo  {journal} {Phys. Rev. Lett.}\ }\textbf
  {\bibinfo {volume} {86}},\ \bibinfo {pages} {3594--3597} (\bibinfo {year}
  {2001})}\BibitemShut {NoStop}%
\bibitem [{\citenamefont {Devakul}\ and\ \citenamefont
  {Huse}(2017)}]{Devakul2017}%
  \BibitemOpen
  \bibfield  {author} {\bibinfo {author} {\bibfnamefont {T.}~\bibnamefont
  {Devakul}}\ and\ \bibinfo {author} {\bibfnamefont {D.~A.}\ \bibnamefont
  {Huse}},\ }\bibfield  {title} {\enquote {\bibinfo {title} {Anderson
  localization transitions with and without random potentials},}\ }\href
  {\doibase 10.1103/PhysRevB.96.214201} {\bibfield  {journal} {\bibinfo
  {journal} {Phys. Rev. B}\ }\textbf {\bibinfo {volume} {96}},\ \bibinfo
  {pages} {214201} (\bibinfo {year} {2017})}\BibitemShut {NoStop}%
\bibitem [{\citenamefont {Deng}\ \emph {et~al.}(2017)\citenamefont {Deng},
  \citenamefont {Ganeshan}, \citenamefont {Li}, \citenamefont {Modak},
  \citenamefont {Mukerjee},\ and\ \citenamefont {Pixley}}]{Deng2017}%
  \BibitemOpen
  \bibfield  {author} {\bibinfo {author} {\bibfnamefont {D.-L.}\ \bibnamefont
  {Deng}}, \bibinfo {author} {\bibfnamefont {S.}~\bibnamefont {Ganeshan}},
  \bibinfo {author} {\bibfnamefont {X.}~\bibnamefont {Li}}, \bibinfo {author}
  {\bibfnamefont {R.}~\bibnamefont {Modak}}, \bibinfo {author} {\bibfnamefont
  {S.}~\bibnamefont {Mukerjee}}, \ and\ \bibinfo {author} {\bibfnamefont
  {J.~H.}\ \bibnamefont {Pixley}},\ }\bibfield  {title} {\enquote {\bibinfo
  {title} {Many-body localization in incommensurate models with a mobility
  edge},}\ }\href {\doibase 10.1002/andp.201600399} {\bibfield  {journal}
  {\bibinfo  {journal} {Annalen der Physik}\ }\textbf {\bibinfo {volume}
  {529}},\ \bibinfo {pages} {1600399} (\bibinfo {year} {2017})}\BibitemShut
  {NoStop}%
\bibitem [{\citenamefont {Ganeshan}\ \emph {et~al.}(2015)\citenamefont
  {Ganeshan}, \citenamefont {Pixley},\ and\ \citenamefont
  {Das~Sarma}}]{Ganeshan2015}%
  \BibitemOpen
  \bibfield  {author} {\bibinfo {author} {\bibfnamefont {S.}~\bibnamefont
  {Ganeshan}}, \bibinfo {author} {\bibfnamefont {J.~H.}\ \bibnamefont
  {Pixley}}, \ and\ \bibinfo {author} {\bibfnamefont {S.}~\bibnamefont
  {Das~Sarma}},\ }\bibfield  {title} {\enquote {\bibinfo {title} {Nearest
  neighbor tight binding models with an exact mobility edge in one
  dimension},}\ }\href {\doibase 10.1103/PhysRevLett.114.146601} {\bibfield
  {journal} {\bibinfo  {journal} {Phys. Rev. Lett.}\ }\textbf {\bibinfo
  {volume} {114}},\ \bibinfo {pages} {146601} (\bibinfo {year}
  {2015})}\BibitemShut {NoStop}%
\bibitem [{\citenamefont {{Kantelhardt}}(2008)}]{Kantelhardt2008}%
  \BibitemOpen
  \bibfield  {author} {\bibinfo {author} {\bibfnamefont {J.~W.}\ \bibnamefont
  {{Kantelhardt}}},\ }\bibfield  {title} {\enquote {\bibinfo {title} {{Fractal
  and Multifractal Time Series}},}\ }\href@noop {} {\bibfield  {journal}
  {\bibinfo  {journal} {ArXiv e-prints}\ } (\bibinfo {year} {2008})},\ \Eprint
  {http://arxiv.org/abs/0804.0747} {arXiv:0804.0747 [physics.data-an]}
  \BibitemShut {NoStop}%
\bibitem [{\citenamefont {Tel}(2014)}]{Tel2014}%
  \BibitemOpen
  \bibfield  {author} {\bibinfo {author} {\bibfnamefont {T.}~\bibnamefont
  {Tel}},\ }\bibfield  {title} {\enquote {\bibinfo {title} {Fractals,
  multifractals, and thermodynamics},}\ }\href {\doibase 10.1515/zna-1988-1221}
  {\bibfield  {journal} {\bibinfo  {journal} {Zeitschrift fur Naturforschung
  A}\ }\textbf {\bibinfo {volume} {43}},\ \bibinfo {pages} {1154--1174}
  (\bibinfo {year} {2014})}\BibitemShut {NoStop}%
\bibitem [{\citenamefont {Edwards}\ and\ \citenamefont
  {Thouless}(1972)}]{Edwards1972}%
  \BibitemOpen
  \bibfield  {author} {\bibinfo {author} {\bibfnamefont {J.~T.}\ \bibnamefont
  {Edwards}}\ and\ \bibinfo {author} {\bibfnamefont {D.~J.}\ \bibnamefont
  {Thouless}},\ }\bibfield  {title} {\enquote {\bibinfo {title} {Numerical
  studies of localization in disordered systems},}\ }\href
  {http://stacks.iop.org/0022-3719/5/i=8/a=007} {\bibfield  {journal} {\bibinfo
   {journal} {Journal of Physics C: Solid State Physics}\ }\textbf {\bibinfo
  {volume} {5}},\ \bibinfo {pages} {807} (\bibinfo {year} {1972})}\BibitemShut
  {NoStop}%
\bibitem [{\citenamefont {Thouless}(1974)}]{Thouless1974}%
  \BibitemOpen
  \bibfield  {author} {\bibinfo {author} {\bibfnamefont {D.~J.}\ \bibnamefont
  {Thouless}},\ }\bibfield  {title} {\enquote {\bibinfo {title} {Electrons in
  disordered systems and the theory of localization},}\ }\href {\doibase
  https://doi.org/10.1016/0370-1573(74)90029-5} {\bibfield  {journal} {\bibinfo
   {journal} {Physics Reports}\ }\textbf {\bibinfo {volume} {13}},\ \bibinfo
  {pages} {93 -- 142} (\bibinfo {year} {1974})}\BibitemShut {NoStop}%
\bibitem [{\citenamefont {Anderson}\ and\ \citenamefont
  {Lee}(1980)}]{Anderson1980_1}%
  \BibitemOpen
  \bibfield  {author} {\bibinfo {author} {\bibfnamefont {P.~W.}\ \bibnamefont
  {Anderson}}\ and\ \bibinfo {author} {\bibfnamefont {P.~A.}\ \bibnamefont
  {Lee}},\ }\bibfield  {title} {\enquote {\bibinfo {title} {The thouless
  conjecture for a one-dimensional chain},}\ }\href {\doibase
  10.1143/PTP.69.212} {\bibfield  {journal} {\bibinfo  {journal} {Progress of
  Theoretical Physics Supplement}\ }\textbf {\bibinfo {volume} {69}},\ \bibinfo
  {pages} {212--219} (\bibinfo {year} {1980})}\BibitemShut {NoStop}%
\bibitem [{\citenamefont {Braun}\ \emph {et~al.}(1997)\citenamefont {Braun},
  \citenamefont {Hofstetter}, \citenamefont {MacKinnon},\ and\ \citenamefont
  {Montambaux}}]{Braun1997}%
  \BibitemOpen
  \bibfield  {author} {\bibinfo {author} {\bibfnamefont {D.}~\bibnamefont
  {Braun}}, \bibinfo {author} {\bibfnamefont {E.}~\bibnamefont {Hofstetter}},
  \bibinfo {author} {\bibfnamefont {A.}~\bibnamefont {MacKinnon}}, \ and\
  \bibinfo {author} {\bibfnamefont {G.}~\bibnamefont {Montambaux}},\ }\bibfield
   {title} {\enquote {\bibinfo {title} {Level curvatures and conductances: A
  numerical study of the thouless relation},}\ }\href {\doibase
  10.1103/PhysRevB.55.7557} {\bibfield  {journal} {\bibinfo  {journal} {Phys.
  Rev. B}\ }\textbf {\bibinfo {volume} {55}},\ \bibinfo {pages} {7557--7564}
  (\bibinfo {year} {1997})}\BibitemShut {NoStop}%
\bibitem [{\citenamefont {Goldberger}\ \emph {et~al.}(2000)\citenamefont
  {Goldberger}, \citenamefont {Amaral}, \citenamefont {Glass}, \citenamefont
  {Hausdorff}, \citenamefont {Ivanov}, \citenamefont {Mark}, \citenamefont
  {Mietus}, \citenamefont {Moody}, \citenamefont {Peng},\ and\ \citenamefont
  {Stanley}}]{Goldberger2000}%
  \BibitemOpen
  \bibfield  {author} {\bibinfo {author} {\bibfnamefont {A.~L.}\ \bibnamefont
  {Goldberger}}, \bibinfo {author} {\bibfnamefont {L.~A.~N.}\ \bibnamefont
  {Amaral}}, \bibinfo {author} {\bibfnamefont {L.}~\bibnamefont {Glass}},
  \bibinfo {author} {\bibfnamefont {J.~M.}\ \bibnamefont {Hausdorff}}, \bibinfo
  {author} {\bibfnamefont {P.~Ch.}\ \bibnamefont {Ivanov}}, \bibinfo {author}
  {\bibfnamefont {R.~G.}\ \bibnamefont {Mark}}, \bibinfo {author}
  {\bibfnamefont {J.~E.}\ \bibnamefont {Mietus}}, \bibinfo {author}
  {\bibfnamefont {G.~B.}\ \bibnamefont {Moody}}, \bibinfo {author}
  {\bibfnamefont {C.-K.}\ \bibnamefont {Peng}}, \ and\ \bibinfo {author}
  {\bibfnamefont {H.~E.}\ \bibnamefont {Stanley}},\ }\bibfield  {title}
  {\enquote {\bibinfo {title} {{PhysioBank, PhysioToolkit, and PhysioNet:
  Components of a New Research Resource for Complex Physiologic Signals}},}\
  }\href {\doibase
  https://www.ahajournals.org/doi/full/10.1161/01.cir.101.23.e215} {\bibfield
  {journal} {\bibinfo  {journal} {Circulation}\ }\textbf {\bibinfo {volume}
  {101}},\ \bibinfo {pages} {e215-- e220} (\bibinfo {year} {2000})}\BibitemShut
  {NoStop}%
\bibitem [{\citenamefont {Dominguez}\ \emph {et~al.}(1992)\citenamefont
  {Dominguez}, \citenamefont {Wiecko},\ and\ \citenamefont
  {Jose}}]{Dominguez1992}%
  \BibitemOpen
  \bibfield  {author} {\bibinfo {author} {\bibfnamefont {D.}~\bibnamefont
  {Dominguez}}, \bibinfo {author} {\bibfnamefont {C.}~\bibnamefont {Wiecko}}, \
  and\ \bibinfo {author} {\bibfnamefont {J.~V.}\ \bibnamefont {Jose}},\
  }\bibfield  {title} {\enquote {\bibinfo {title} {{Wave-function and
  resistance scaling for quadratic irrationals in Harper's equation}},}\
  }\href@noop {} {\bibfield  {journal} {\bibinfo  {journal} {Phys. Rev. B}\
  }\textbf {\bibinfo {volume} {45}},\ \bibinfo {pages} {13919} (\bibinfo {year}
  {1992})}\BibitemShut {NoStop}%
\bibitem [{\citenamefont {{Mac\'e, Nicolas and Jagannathan, Anuradha and
  Pi\'echon, Fr\'ed\'eric}}(2016)}]{Mace2016}%
  \BibitemOpen
  \bibfield  {author} {\bibinfo {author} {\bibnamefont {{Mac\'e, Nicolas and
  Jagannathan, Anuradha and Pi\'echon, Fr\'ed\'eric}}},\ }\bibfield  {title}
  {\enquote {\bibinfo {title} {Fractal dimensions of wave functions and local
  spectral measures on the fibonacci chain},}\ }\href {\doibase
  10.1103/PhysRevB.93.205153} {\bibfield  {journal} {\bibinfo  {journal} {Phys.
  Rev. B}\ }\textbf {\bibinfo {volume} {93}},\ \bibinfo {pages} {205153}
  (\bibinfo {year} {2016})}\BibitemShut {NoStop}%
\bibitem [{\citenamefont {{Mac\'e, Nicolas and Jagannathan, Anuradha and
  Kalugin, Pavel and Mosseri, R\'emy and Pi\'echon,
  Fr\'ed\'eric}}(2017)}]{Mace2017}%
  \BibitemOpen
  \bibfield  {author} {\bibinfo {author} {\bibnamefont {{Mac\'e, Nicolas and
  Jagannathan, Anuradha and Kalugin, Pavel and Mosseri, R\'emy and Pi\'echon,
  Fr\'ed\'eric}}},\ }\bibfield  {title} {\enquote {\bibinfo {title} {Critical
  eigenstates and their properties in one- and two-dimensional
  quasicrystals},}\ }\href {\doibase 10.1103/PhysRevB.96.045138} {\bibfield
  {journal} {\bibinfo  {journal} {Phys. Rev. B}\ }\textbf {\bibinfo {volume}
  {96}},\ \bibinfo {pages} {045138} (\bibinfo {year} {2017})}\BibitemShut
  {NoStop}%
\bibitem [{\citenamefont {Landauer}(1970)}]{Landauer1970}%
  \BibitemOpen
  \bibfield  {author} {\bibinfo {author} {\bibfnamefont {R.}~\bibnamefont
  {Landauer}},\ }\bibfield  {title} {\enquote {\bibinfo {title} {Electrical
  resistance of disordered one-dimensional lattices},}\ }\href {\doibase
  10.1080/14786437008238472} {\bibfield  {journal} {\bibinfo  {journal} {The
  Philosophical Magazine: A Journal of Theoretical Experimental and Applied
  Physics}\ }\textbf {\bibinfo {volume} {21}},\ \bibinfo {pages} {863--867}
  (\bibinfo {year} {1970})}\BibitemShut {NoStop}%
\bibitem [{\citenamefont {Economou}\ and\ \citenamefont
  {Soukoulis}(1981)}]{Economou1981}%
  \BibitemOpen
  \bibfield  {author} {\bibinfo {author} {\bibfnamefont {E.~N.}\ \bibnamefont
  {Economou}}\ and\ \bibinfo {author} {\bibfnamefont {C.~M.}\ \bibnamefont
  {Soukoulis}},\ }\bibfield  {title} {\enquote {\bibinfo {title} {Static
  conductance and scaling theory of localization in one dimension},}\ }\href
  {\doibase 10.1103/PhysRevLett.46.618} {\bibfield  {journal} {\bibinfo
  {journal} {Phys. Rev. Lett.}\ }\textbf {\bibinfo {volume} {46}},\ \bibinfo
  {pages} {618--621} (\bibinfo {year} {1981})}\BibitemShut {NoStop}%
\bibitem [{\citenamefont {Fisher}\ and\ \citenamefont
  {Lee}(1981)}]{Fisher1981}%
  \BibitemOpen
  \bibfield  {author} {\bibinfo {author} {\bibfnamefont {D.~S.}\ \bibnamefont
  {Fisher}}\ and\ \bibinfo {author} {\bibfnamefont {P.~A.}\ \bibnamefont
  {Lee}},\ }\bibfield  {title} {\enquote {\bibinfo {title} {Relation between
  conductivity and transmission matrix},}\ }\href {\doibase
  10.1103/PhysRevB.23.6851} {\bibfield  {journal} {\bibinfo  {journal} {Phys.
  Rev. B}\ }\textbf {\bibinfo {volume} {23}},\ \bibinfo {pages} {6851--6854}
  (\bibinfo {year} {1981})}\BibitemShut {NoStop}%
\bibitem [{\citenamefont {L\"uschen}\ \emph {et~al.}(2017)\citenamefont
  {L\"uschen}, \citenamefont {Bordia}, \citenamefont {Scherg}, \citenamefont
  {Alet}, \citenamefont {Altman}, \citenamefont {Schneider},\ and\
  \citenamefont {Bloch}}]{Luschen2017}%
  \BibitemOpen
  \bibfield  {author} {\bibinfo {author} {\bibfnamefont {H.~P.}\ \bibnamefont
  {L\"uschen}}, \bibinfo {author} {\bibfnamefont {P.}~\bibnamefont {Bordia}},
  \bibinfo {author} {\bibfnamefont {S.}~\bibnamefont {Scherg}}, \bibinfo
  {author} {\bibfnamefont {F.}~\bibnamefont {Alet}}, \bibinfo {author}
  {\bibfnamefont {E.}~\bibnamefont {Altman}}, \bibinfo {author} {\bibfnamefont
  {U.}~\bibnamefont {Schneider}}, \ and\ \bibinfo {author} {\bibfnamefont
  {I.}~\bibnamefont {Bloch}},\ }\bibfield  {title} {\enquote {\bibinfo {title}
  {Observation of slow dynamics near the many-body localization transition in
  one-dimensional quasiperiodic systems},}\ }\href {\doibase
  10.1103/PhysRevLett.119.260401} {\bibfield  {journal} {\bibinfo  {journal}
  {Phys. Rev. Lett.}\ }\textbf {\bibinfo {volume} {119}},\ \bibinfo {pages}
  {260401} (\bibinfo {year} {2017})}\BibitemShut {NoStop}%
\bibitem [{\citenamefont {Vosk}\ \emph {et~al.}(2015)\citenamefont {Vosk},
  \citenamefont {Huse},\ and\ \citenamefont {Altman}}]{Vosk2015}%
  \BibitemOpen
  \bibfield  {author} {\bibinfo {author} {\bibfnamefont {R.}~\bibnamefont
  {Vosk}}, \bibinfo {author} {\bibfnamefont {D.~A.}\ \bibnamefont {Huse}}, \
  and\ \bibinfo {author} {\bibfnamefont {E.}~\bibnamefont {Altman}},\
  }\bibfield  {title} {\enquote {\bibinfo {title} {Theory of the many-body
  localization transition in one-dimensional systems},}\ }\href {\doibase
  10.1103/PhysRevX.5.031032} {\bibfield  {journal} {\bibinfo  {journal} {Phys.
  Rev. X}\ }\textbf {\bibinfo {volume} {5}},\ \bibinfo {pages} {031032}
  (\bibinfo {year} {2015})}\BibitemShut {NoStop}%
\bibitem [{\citenamefont {Potter}\ \emph {et~al.}(2015)\citenamefont {Potter},
  \citenamefont {Vasseur},\ and\ \citenamefont {Parameswaran}}]{Potter2015}%
  \BibitemOpen
  \bibfield  {author} {\bibinfo {author} {\bibfnamefont {A.~C.}\ \bibnamefont
  {Potter}}, \bibinfo {author} {\bibfnamefont {R.}~\bibnamefont {Vasseur}}, \
  and\ \bibinfo {author} {\bibfnamefont {S.~A.}\ \bibnamefont {Parameswaran}},\
  }\bibfield  {title} {\enquote {\bibinfo {title} {Universal properties of
  many-body delocalization transitions},}\ }\href {\doibase
  10.1103/PhysRevX.5.031033} {\bibfield  {journal} {\bibinfo  {journal} {Phys.
  Rev. X}\ }\textbf {\bibinfo {volume} {5}},\ \bibinfo {pages} {031033}
  (\bibinfo {year} {2015})}\BibitemShut {NoStop}%
\bibitem [{\citenamefont {{Zhang}}\ and\ \citenamefont
  {{Yao}}(2018)}]{Zhang2018}%
  \BibitemOpen
  \bibfield  {author} {\bibinfo {author} {\bibfnamefont {S.-X.}\ \bibnamefont
  {{Zhang}}}\ and\ \bibinfo {author} {\bibfnamefont {H.}~\bibnamefont
  {{Yao}}},\ }\bibfield  {title} {\enquote {\bibinfo {title} {{Universal
  properties of many-body localization transitions in quasiperiodic
  systems}},}\ }\href@noop {} {\bibfield  {journal} {\bibinfo  {journal} {ArXiv
  e-prints}\ } (\bibinfo {year} {2018})},\ \Eprint
  {http://arxiv.org/abs/1805.05958} {arXiv:1805.05958 [cond-mat.str-el]}
  \BibitemShut {NoStop}%
\bibitem [{\citenamefont {Akkermans}(1997)}]{Akkermans1997}%
  \BibitemOpen
  \bibfield  {author} {\bibinfo {author} {\bibfnamefont {E.}~\bibnamefont
  {Akkermans}},\ }\bibfield  {title} {\enquote {\bibinfo {title} {Twisted
  boundary conditions and transport in disordered systems},}\ }\href {\doibase
  10.1063/1.531913} {\bibfield  {journal} {\bibinfo  {journal} {Journal of
  Mathematical Physics}\ }\textbf {\bibinfo {volume} {38}},\ \bibinfo {pages}
  {1781--1793} (\bibinfo {year} {1997})}\BibitemShut {NoStop}%
\bibitem [{\citenamefont {Zhou}\ \emph {et~al.}(2013)\citenamefont {Zhou},
  \citenamefont {Dang},\ and\ \citenamefont {R.}}]{Zhou2013}%
  \BibitemOpen
  \bibfield  {author} {\bibinfo {author} {\bibfnamefont {W.}~\bibnamefont
  {Zhou}}, \bibinfo {author} {\bibfnamefont {Y.}~\bibnamefont {Dang}}, \ and\
  \bibinfo {author} {\bibfnamefont {Gu}~\bibnamefont {R.}},\ }\bibfield
  {title} {\enquote {\bibinfo {title} {{Efficiency and multifractality analysis
  of CSI 300 based on multifractal detrending moving average algorithm}},}\
  }\href@noop {} {\bibfield  {journal} {\bibinfo  {journal} {Physica A}\
  }\textbf {\bibinfo {volume} {392}},\ \bibinfo {pages} {1429--1438} (\bibinfo
  {year} {2013})}\BibitemShut {NoStop}%
\bibitem [{\citenamefont {Markos}(2006)}]{Markos2006}%
  \BibitemOpen
  \bibfield  {author} {\bibinfo {author} {\bibfnamefont {P.}~\bibnamefont
  {Markos}},\ }\bibfield  {title} {\enquote {\bibinfo {title} {{Numerical
  Analysis of The Anderson Localization}},}\ }\href@noop {} {\bibfield
  {journal} {\bibinfo  {journal} {Acta Physica Slovaca}\ }\textbf {\bibinfo
  {volume} {56}},\ \bibinfo {pages} {561--686} (\bibinfo {year}
  {2006})}\BibitemShut {NoStop}%
\bibitem [{\citenamefont {Verges}(1999)}]{Verges1999}%
  \BibitemOpen
  \bibfield  {author} {\bibinfo {author} {\bibfnamefont {J.~A.}\ \bibnamefont
  {Verges}},\ }\bibfield  {title} {\enquote {\bibinfo {title} {{Computational
  implementation of the Kubo formula for the static conductance: application to
  two-dimensional quantum dots}},}\ }\href@noop {} {\bibfield  {journal}
  {\bibinfo  {journal} {Computer Physics Communications}\ }\textbf {\bibinfo
  {volume} {118}},\ \bibinfo {pages} {71--80} (\bibinfo {year}
  {1999})}\BibitemShut {NoStop}%
\end{thebibliography}%

\begin{widetext}
\renewcommand{\thesection}{S\arabic{section}}    
\renewcommand{\thefigure}{S\arabic{figure}}
\renewcommand{\theequation}{S\arabic{equation}} 


\setcounter{figure}{0}
\setcounter{equation}{0}
\setcounter{section}{0}

\section{Supplementary Information}

\subsection{Thouless conductance} \label{SIsec:Thouless}
The Thouless conductance, discussed in Sec.\ref{sec:Thouless} of the main text, is defined as
\begin{align}
\text{g}_{T}(E)&=\frac{\delta E}{\Delta_E}\label{SIeq:Thouless}
\end{align}
where $\Delta_E$ is the mean level spacing and $\delta E$ is the geometric mean of energy level shifts, $|\epsilon_A-\epsilon_P|$, over an energy window $[E-w,E+w]$ with width $w\gg \Delta_E$. Here $\epsilon_{P}$ and $\epsilon_{A}$ are eigenvalues of the Hamiltonian with periodic and anti-periodic boundary conditions, respectively. We calculate the energy spectrum by numerical diagonalization of the quasiperiodic Hamiltonians considered in the main text. The energy spectrum of the Aubry-Andre model has a Cantor set structure with bands of states separated by dense set of gaps \cite{Simon1982,Ostlund1984,Sokoloff1985}. We choose $w$ to be much smaller than the width of the principal bands. Alternatively, $g_\mr{T}$ can be defined in terms of the mean energy level curvature under a twisted boundary condition or an Aharonov-Bohm flux in a ring geometry \cite{Akkermans1997,Braun1997}. We have checked that $g_\mr{T}$ obtained from mean energy level curvature gives results similar to that in Eq.\eqref{SIeq:Thouless}. Since the latter does not require the computation of eigenvectors, we have used Eq.\eqref{SIeq:Thouless} to calculate $g_\mr{T}$, reported in the main text. We obtain the mean, $\langle g_\mr{T}\rangle$, and the typical, $\langle \exp{(g_\mr{T})}\rangle$, conductances by averaging over $\phi$. We also calculate $g_\mr{T}^\infty$, averaged over the whole energy spectrum, as shown in Fig.\ref{SIfig:Thouless}(a) for the critical point ($V=1$) in $1d$. This shows the sharp resonances and various sequences of lengths with different powerlaws, as in Fig.\ref{fig:Conductance}(b) (main text) for $g_\mr{T}(E=0)$. 
 
We have also calculated $g_\mr{T}(L)$ in $1d$ for the irrational number $b=1/\sigma_s=\sqrt{2}-1$, the reciprocal of silver ratio. As shown in Fig.\ref{SIfig:Thouless}(b), here also we get similar peaks in the conductance at system sizes related to the Pell numbers, i.e. $P_{n+1}=2P_n +P_{n-1}$, with $P_0=1$ and $P_1=2$, such that $\sigma_s= \lim_{n\to \infty} (P_{n+1}/P_n)$.
 
\begin{figure}[h!]
\begin{center}
\includegraphics[width=0.6\linewidth]{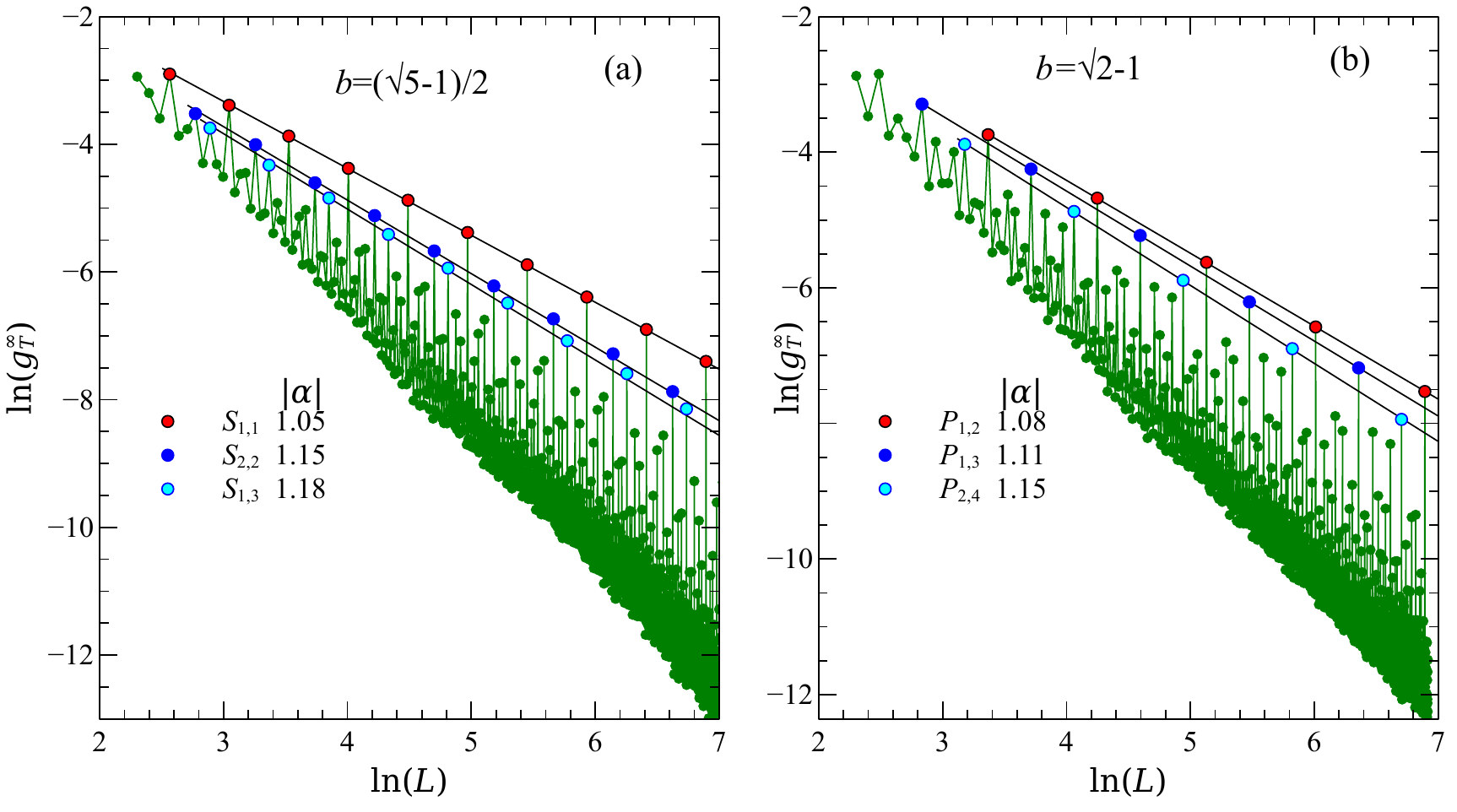}
\caption{(a) The infinite temperature (i.e. averaged over the full energy spectrum) Thouless conductance ($g^{\infty}_T$) for $1d$, with the inverse of golden ratio, $(\sqrt{5}-1)/2$ as the irrational number ($b$) in the potential, is shown varying the system size. $\mathcal{S}$'s are the same sequences described in Fig.\ref{fig:Conductance}. Conductance for different sequence of system size varies with different exponents $\alpha$, where  $g_\mr{T}\propto L^{\alpha}$. (b) Now the irrational number is changed to the inverse of silver ratio, $(\sqrt{2}-1)$ and $g_T^{\infty}$ is calculated. $P_{L_1,L_2}$ represents the sequence of system sizes (related to Pell numbers) started with seeds $L_1$ and $L_2$ (see \ref{SIsec:Thouless}). Both in (a) and (b), the peak of the conductance appears when the system size belongs to the respective sequences.}
  \label{SIfig:Thouless}
\end{center}
\end{figure}

\subsection{Multifractal Analysis} \label{SIsec:Multifractal}
Motivated by the strong-non monotonicity of $g_\mr{T}(L)$ in $1d$ [Figs.\ref{fig:Conductance}(a),(b), main text] and multiple powerlaws in Fig.\ref{fig:Conductance}(b), we carry out a multifractal fluctuation analysis \cite{Kantelhardt2008,Zhou2013} of the $g_\mr{T}(L)$ data, treating it as a \emph{time series}, i.e. $g_\mr{T}(i)\equiv g_\mr{T}(L_i)$ with $i=1,\dots,N$, where $L_1$ and $L_N$ are minimum and maximum system sizes studied, respectively. First, we do a cumulative sum of the data, i.e. $y(j)=\sum_{i=1}^j g_\mr{T}(i)$ for $j=1,\dots,N$. Then, to remove any trend from the data, we subtract moving average from each data point. The moving average $\bar{y}(j)$ is the average of $y(j)$'s over an interval (here we used $[j-13,j+13]$) around $j$. This gives us the residual sequence $\tilde{y}(j)=y(j)-\bar{y}(j)$. Now the residual sequence is divide into non-overlapping segments $j_s=1,\dots,N_s$ of width $s$, where $N_s$ is the largest integer not larger than $N/s-1$. The root mean square (rms) fluctuation is calculated for each segment, i.e.
\begin{subequations}\label{SIeq:MultifractalMoments}
\begin{align}
F_s(j_s)&=\sqrt{\frac{1}{s}\sum_{j\in j_s}\bar{y}^2(j)}
\end{align}
and various moments, i.e.
\begin{align}
P_q&=\left(\frac{1}{N_s}\sum_{j_s=1}^{N_s} F_s^q(j_s)\right)^{1/q}
\end{align}
\end{subequations}
are calculated. These moments follow multifractal powelaw scalings with the segments length $s$, i.e. $P_q(s)\sim s^{h(q)}$, as shown for $q=3,4$ in Fig.\ref{fig:Conductance}(e) (main text) at the $1d$ critical point. 

To quantify the multifractality, we can obtain the singularity spectrum through a Legendre transform, $f(\alpha)=q[\alpha -h(q)]+1$, with $\alpha=\partial[qh(q)-1]/\partial q$. For a more refined multifractal analysis, we use a wavelet transform the Thouless conductance data, namely we convolve the data set with a fixed order derivative of the Gaussian funtion, ${G}^n(x)= d^n (e^{-x^2/2})/dx^n $. This removes any polynomial trend in the data upto order $n-1$ leaving only the singular dependence. Now this power can be extracted via a log-log fitting and thus, the singularity spectrum can be obtained. To this end, we use the codes of ref.\cite{Goldberger2000}. In our calculation we use the fourth order derivative of Gaussian function. The resulting singularity spectra $f(\alpha)$ for $1d$ Thouless conductance in the metallic phase and at the critical point are shown in Fig.\ref{fig:Conductance}(f). The singularity spectra computed using moments in Eqs.\ref{SIeq:MultifractalMoments} are qualitatively similar to that obtained via wavelet transform method.

\subsubsection{Wavefunction Multifractality}
The conductance multifractality obtained in the preceding section from the $L$ dependence of Thouless conductance directly characterizes the violation of the assumption of monotonicity in single-parameter scaling theory. As discussed in the main text, this kind of multifractality in quasiperiodic system is quite different from well-known the wavefunction multifractality at the critical point between metal and insulator in random system, e.g. at the 3d Anderson transition \cite{Evers2008}.  Conventionally, the multifractality of critical single-particle eigenstates $\psi_r$ is analyzed in terms of the moments of wavefunction amplitude \cite{Evers2008}, i.e. $P^{\psi}_q=\sum_r |\psi_r|^{2q}$, which upon disorder averaging follows a powerlaw scaling $\langle P_q\rangle\sim L^{-\tau(q)}$, with an exponent $\tau(q)=d(q-1)+\Delta_q$ non-trivially dependent on $q$, as characterized by the anomalous dimension $\Delta(q)$. We show in Fig.\ref{SIfig:1D_QP_wv_fn_mf} that, much like at the $3d$ Anderson criticality \cite{Evers2008}, the critiacal wavefunctions of quasiperiodic system also possess the usual multifractality in $1d$ as characterized by the singularity spectrum obtained from the Legendre transform of $\tau(q)$ \cite{Dominguez1992}.
\begin{figure*}[htb!]
\centering
\includegraphics[width=0.8\textwidth]{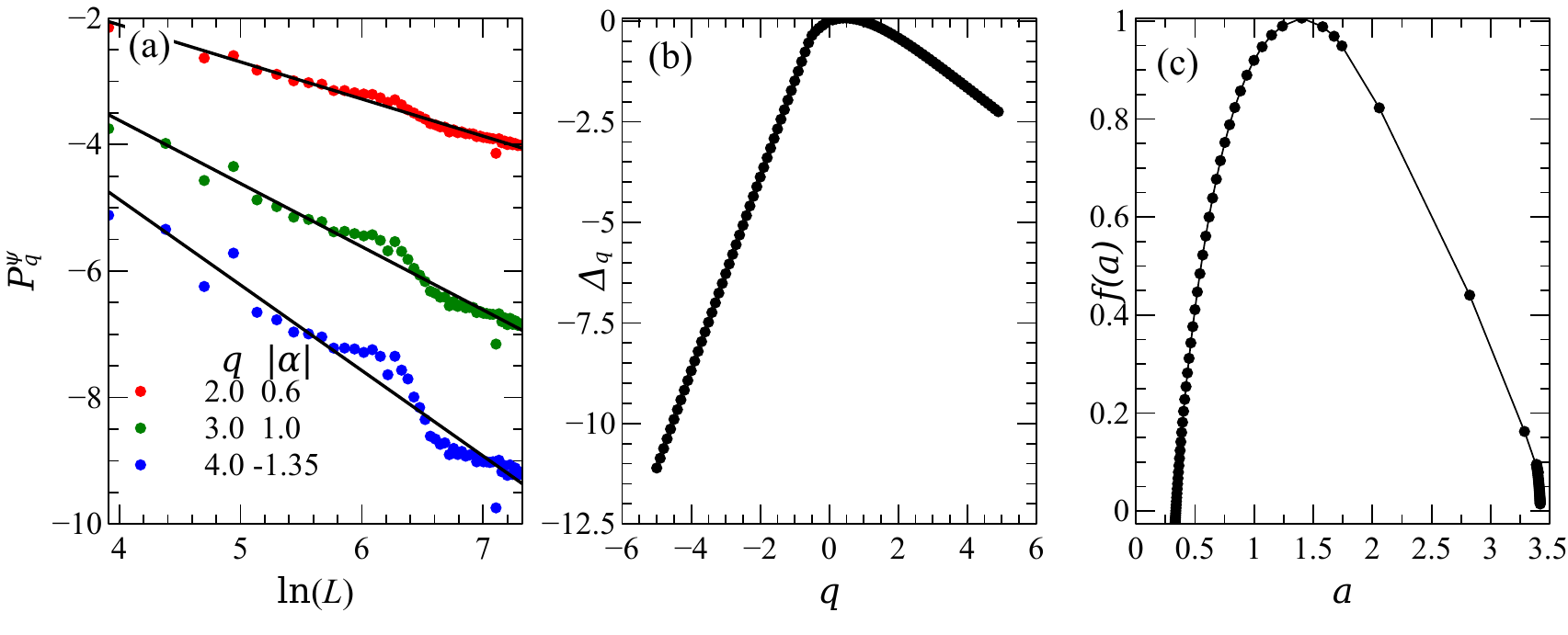}
\caption{(a) The scaling of different moments of the wave function of the 1D Hamiltonian, $P_q^{\psi}\sim L^{-\alpha}$, averaged over all energy, is shown here. (b) Shows the anomalous dimension. (c) The singularity spectrum shows the multifractal nature.}
\label{SIfig:1D_QP_wv_fn_mf}
\end{figure*}

\subsection{Landauer conductance in $1d$}
Schrodinger equation for the $1d$ Hamiltonian, given in Eq.\eqref{eq.AAmodel}, can be written in the latttice basis, $\{\psi_r\}$ in the following way
\begin{equation}
\left( {\begin{array}{c}
   \psi_{r+1} \\
   \psi_{r}  \\
  \end{array} }\right)
   =
   \left( {\begin{array}{cc}
   \epsilon_r & -1 \\
   1 & 0 \\
  \end{array} } \right)
  \left( {\begin{array}{c}
   \psi_r \\
   \psi_{r-1} \\
  \end{array} } \right)
  =M_r\left( {\begin{array}{c}
   \psi_{r} \\
   \psi_{r-1}  \\
  \end{array} }\right)
  =\prod_{i=1}^r M_i
    \left( {\begin{array}{c}
    \psi_{1} \\
   \psi_{0}  \\
  \end{array} }\right)= \textbf{M}
    \left( {\begin{array}{c}
    \psi_{1} \\
   \psi_{0}  \\
  \end{array} }\right)    ,
\end{equation}
where $\epsilon_r= E-2V \cos(2\pi b r+\phi)$ and $E$ is the energy of interest. Iterating this equation we can calculate the amplitude at the end points given the two starting amplitudes. This transfer matrix $\textbf{M}$ is related to the transmission matrix $\textbf{T}$ via a transformation \cite{Markos2006},
\begin{equation}
\textbf{T}=\textbf{Q}^{-1}\textbf{MQ},
\end{equation} 
where 
\begin{equation}
\textbf{Q}=
 \left( {\begin{array}{cc}
   1 & 1 \\
   e^{-ik} & e^{ik} \\
  \end{array} } \right).
\end{equation}
Here the disordered region is considered to exist for $N>i>0$ and $V=0$ at all other points, with transmitted wave amplitude $\psi_{-1}=e^{ik}$ and $\psi_0=1$ for a wave propagating from $i>N$ region to $i<0$.
The Landauer conductance is given by
\begin{equation}
g_\mr{L}= \dfrac{|t|^2}{|r|^2},
\end{equation} 
where $t$ and $r$ is the transmission and reflection amplitude in the transmission matrix. The Landauer conductance $g_\mr{L}(L)$ is shown in Figs.\ref{fig:Disordered_landauer_cond}(a)-(d) for metallic and critical states. The strong non-monotonicty in $g_\mr{L}(L)$ is evident.

\begin{figure*}[htb!]
\centering
\includegraphics[width=0.8\textwidth]{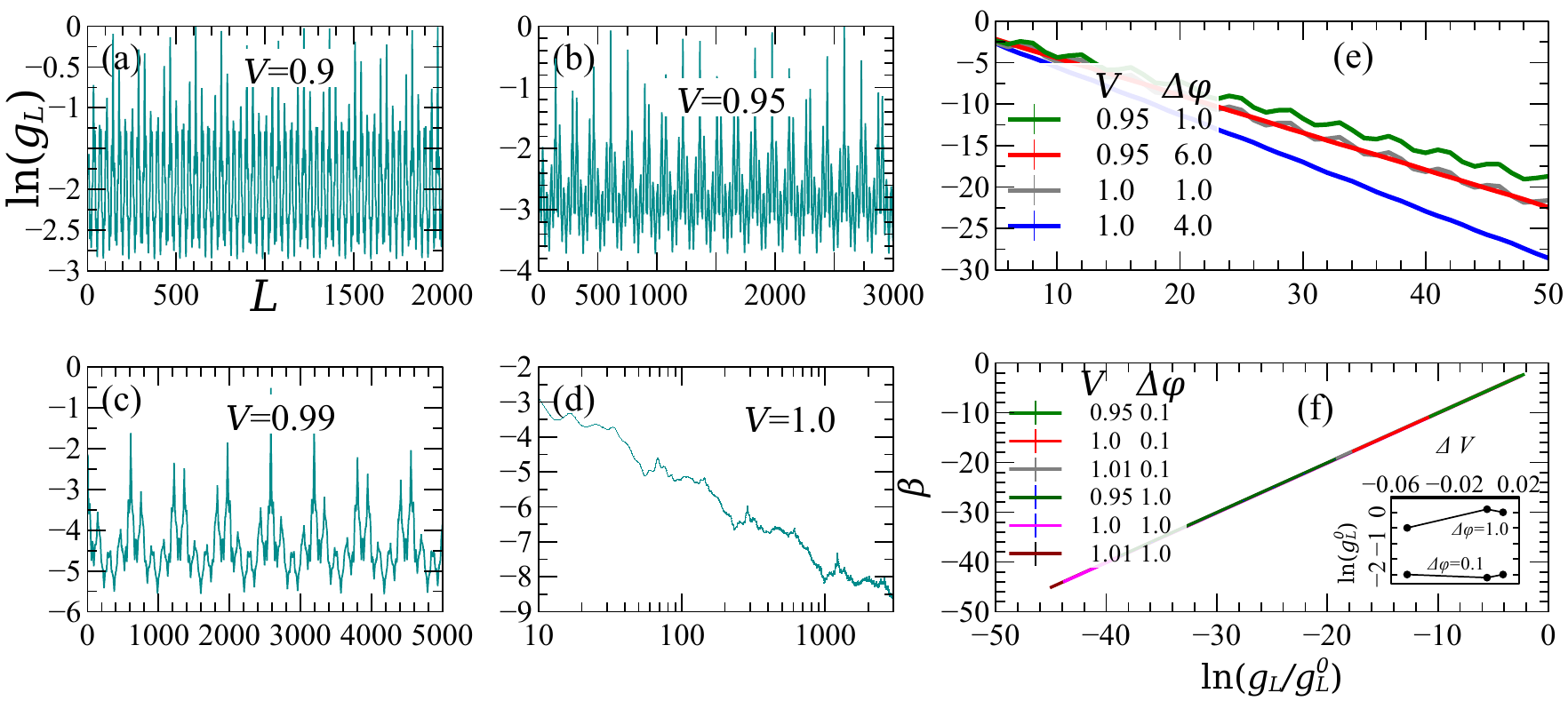}
\caption{(b)-(e) have the same axes labeling as (a). (a)-(d) $g_\mr{L}(L)$ for $1d$, both in metallic side and at the critical point, are shown. (e) In the presence of randomly chosen phase at each site, from a uniform random disorder $[-\Delta \phi/2,\Delta \phi/2]$, the system becomes insulator for any nonzero value of $V$ and $\Delta \phi$, which indicates the perturbation to be relevant.For a weak phase disorder the monotonicity of $g_\mr{L}(L)$ is still present. The data points are averaged over $3000$ disorder realization. (f) Ignoring the $L$ dependent fluctuation in weak disorder a continuous $\beta$ function is obtained, hence the single-parameter scaling theory is recovered for the large length scale behavior of the conductance.}
\label{fig:Disordered_landauer_cond}
\end{figure*}

\subsection{Kubo conductance}  
The open-system (dimensionless) conductance at energy $E$ for the system described by the quasiperiodic Hamiltonians [Eqs.\eqref{eq.AAmodel},\eqref{eq:model}] connected with non-interacting leads at the two ends along $x$ direction is given by the Kubo formula \cite{Fisher1981,Lee1981},
\begin{equation}
g_\mr{K} (E)= 2\text{Tr}[\hat{I}^x(x)\hat{G}''(E)\hat{I}^x(x')\hat{G}''(E)].
\end{equation}
Where $G"=(1/2i) (G^--G^+)$ is obtained in terms of the Green's functions $G^\pm(E)=(E-H\pm i\eta)^{-1}$, $H$ being the Hamiltonian of the whole system including the leads. The current operator is
\begin{align}
\hat{I}(j)= it\sum_{l} ( \ket{j-1,l}\bra{j,l} -\ket{j,l}\bra{j-1,l})  
\end{align}
$l$ is the index to represent sites on any slice $j$ perpendicular to the direction $x$. The conductance then simplifies to 
\begin{equation}
g_\mr{K}= 2\text{Tr}[2 G''(j,j)G''(j-1,j-1)-G''(j-1,j)^2-G''(j,j-1)^2].
\label{SIeq:conductance_formula}
\end{equation}
The trace is over $l$ i.e in the transverse direction. We evaluate the conductance by calculating the Green's functions in Eq.\eqref{SIeq:conductance_formula} via the standard recursive Green's function method described in refs.\cite{Lee1981,MacKinnon1983,Verges1999}. The attached leads have same width as that of the system and we use hard-wall or open boundary condition in the transverse directions.

\subsection{Beta function calculations}
To extract the beta functions in Figs.\ref{fig:BetaFunction} (main text), we carry out linear fitting for the $\ln{g}$ vs. $\ln{L}$ curves in the region $V\leq 1$ and, for $V>1$, we do the same for $\ln{g}$ vs. $L$ curves. This gives a powerlaw dependency of conductance on $L$ for metallic phase ($V\leq 1$) and exponential dependency in the insulating regime ($V>1$). The scaling-theory beta function $\beta(g)= d\ln{g}/d\ln{L}$ is calculated by taking logarithmic derivative of the fitting curves. In $3d$, close to the critical point. we perform a scaling collapse of the data following ref.\cite{Slevin2001}. To this end, we assume a single-parameter finite-size scaling form for the conductance, namely 
\begin{equation}
\ln{g} = \mathcal{F}(\Psi L^{1/\nu})
\end{equation}
The relevant scaling variable $\Psi$, in terms of dimensionless parameter $v=(V-V_c)/V_c$, is approximated  as $\Psi= \Psi_1 v + \Psi_2 v^2$ and expand the scaling function $\mathcal{F}$ upto third-order polynomial. We minimize the quantity $\sum_i (\ln{g_i}- \mathcal{F}(\Psi_iL_i^{1/\nu}))^2$ to obtain the fitting parameters, $V_c$, $\nu$, $\Psi_1$, $\Psi_2$ and the coefficients of the third-order polynomial, where index $i$ represents each point of the data set $\{V,L\}$. Once the scaling function $\mathcal{F}(x)$ is known in terms of these parameters, we calculate the smooth $\beta$ function in $3d$ near the metal-insulator transition at $V_c\simeq 2.2$, as shown in Fig.\ref{fig:Conductance_HigherD}(d). 

\subsubsection{Effects of phase disorder}
 In Fig.\ref{fig:Disordered_landauer_cond}(f), the results for $\beta(\tilde{g}_\mr{L})$ is shown for a $1d$ model where we modify the quasiperiodic potential in Eq.\eqref{eq.AAmodel} from $\cos(2\pi b r+\phi)$ to $\cos(2\pi b r+\phi_r)$ with $\phi_r$ an uncorrelated random phase at each site chosen uniformly from $[-\Delta \phi/2,\Delta \phi/2]$. The phase randomness, even if weak, leads to localization, as eveident from exponential decay of conductance with $L$ in Fig.\ref{fig:Disordered_landauer_cond}(e), even for $V<1$. As expected for a random system, one gets back a continuous beta function, considering the conductance dependence on long window of system sizes, i.e. ignoring the non-monotonicity with $L$ at small lengths, for weak strength of the randomness, in contrast to that in Fig.\ref{fig:BetaFunction}(b). As shown in Fig.\ref{fig:Disordered_landauer_cond}(e), the bare $g_\mr{L}(L)$ has a non-monotonic behavior with $L$ in the presence of weak phase randomness, but the non monotonicity goes away as the randomness increases, completely restoring single parameter scaling theory even for moderate strength of disorder.
\end{widetext}
\end{document}